\documentclass[structabstract]{aa}  
\usepackage{graphicx}
\usepackage{txfonts}
\usepackage{natbib}
\usepackage{longtable}
\def\pdeg{\ifmmode $\setbox0=\hbox{$^{\circ}$}\rlap{\hskip.11\wd0 .}$^{\circ}
  \else \setbox0=\hbox{$^{\circ}$}\rlap{\hskip.11\wd0 .}$^{\circ}$\fi}

\begin{document}
\title{OH maser emission in the THOR survey of the northern Milky Way\thanks{The maser
    Table 1 is available in electronic form at the CDS via anonymous ftp to
    cdsarc.u-strasbg.fr (130.79.128.5) or via
    http://cdsweb.u-strasbg.fr/cgi-bin/qcat?J/A+A/}.}


   \author{H.~Beuther
          \inst{1}
          \and
          A.~Walsh
          \inst{2}
           \and
          Y.~Wang
          \inst{1}
          \and
          M.~Rugel
          \inst{1,3}
           \and
          J.~Soler
          \inst{1}
           \and
          H.~Linz
          \inst{1}
           \and
          R.S.~Klessen
          \inst{4}
           \and
          L.D.~Anderson
          \inst{5,10,11}
           \and
          J.S.~Urquhart
          \inst{6}
           \and
          S.C.O.~Glover
          \inst{4}
           \and
          S.J.~Billington
          \inst{6}
           \and
          J.~Kainulainen
          \inst{7,1}
           \and
          K.M.~Menten
          \inst{3}
           \and
          N.~Roy
          \inst{8}
           \and
           S.N.~Longmore
          \inst{9}
          \and
           F.~Bigiel
          \inst{12}
}
   \institute{$^1$ Max Planck Institute for Astronomy, K\"onigstuhl 17,
              69117 Heidelberg, Germany, \email{beuther@mpia.de}\\
              $^2$ Research Centre for Astronomy, Astrophysics, and Astrophotonics, Macquarie University, NSW 2109, Australia\\
              $^3$ Max-Planck-Institut f\"ur Radioastronomie, Auf dem H\"ugel 69, 53121 Bonn, Germany\\
              $^4$ Universit\"at Heidelberg, Zentrum f\"ur Astronomie, Institut f\"ur Theoretische Astrophysik, Albert-Ueberle-Str. 2, D-69120 Heidelberg, Germany\\
              $^5$ Department of Physics and Astronomy, West Virginia University, Morgantown, WV 26506, USA\\
              $^{6}$ School of Physical Sciences, University of Kent, Ingram Building, Canterbury, Kent CT2 7NH, UK\\
              $^{7}$ Chalmers University of Technology, Department of Space, Earth and Environment, SE-412 93 Gothenburg, Sweden\\
              $^{8}$ Department of Physics, Indian Institute of Science, Bangalore 560012, India\\
              $^{9}$ Astrophysics Research Institute, Liverpool John Moores University, 146 Brownlow Hill, Liverpool L3 5RF, UK\\
              $^{10}$ Center for Gravitational Waves and Cosmology, West Virginia University, Chestnut Ridge Research Building, Morgantown, WV 26505, USA\\
              $^{11}$ Adjunct Astronomer at the Green Bank Observatory, P.O. Box 2, Green Bank, WV 24944, USA\\
             $^{12}$ Argelander-Institute for Astronomy, University of Bonn, Auf dem H\"ugel 71, 53121 Bonn, Germany
}

   \date{Version of \today}

\abstract
{OH masers trace diverse physical processes, from the expanding
  envelopes around evolved stars to star-forming regions or supernovae
  remnants. Providing a survey of the ground-state OH maser
  transitions in the northern hemisphere inner Milky Way facilitates
  the study of a broad range of scientific topics.}
{We want to identify the ground-state OH masers at $\sim$18\,cm
  wavelength in the area covered by ``The HI/OH/Recombination line
  survey of the Milky Way (THOR)''. We will present a catalogue of all
  OH maser features and their possible associated environments.}
{The THOR survey covers longitude and latitude ranges of
  $14\pdeg3<l<66\pdeg8$ and $b<\pm 1\pdeg25$. All OH ground state
  lines $^2\Pi_{3/2}$(J=3/2) at 1612 (F=1-2), 1665 (F=1-1), 1667
  (F=2-2) and 1720\,MHz (F=2-1) have been observed, employing the Very
  Large Array (VLA) in its C configuration. The spatial resolution of
  the data varies between $12.5''$ and $19''$, the spectral resolution
  is 1.5\,km\,s$^{-1}$, and the rms sensitivity of the data is
  $\sim$10\,mJy\,beam$^{-1}$ per channel.}
{We identify 1585 individual maser spots (corresponding to single
  spectral features) distributed over 807 maser sites (regions of size
  $\sim 10^3-10^4$\,AU). Based on different criteria from spectral
  profiles to literature comparison, we try to associate the maser
  sites with astrophysical source types. Approximately 51\% of the
  sites exhibit the double-horned 1612\,MHz spectra typically emitted
  from the expanding shells of evolved stars. The separations of the
  two main velocity features of the expanding shells typically vary
  between 22 and 38\,km\,s$^{-1}$. In addition to this, at least 20\%
  of the maser sites are associated with star-forming regions. While
  the largest fraction of 1720\,MHz maser spots (21 out of 53)
  is associated with supernova remnants, a significant fraction of the
  1720\,MHz maser spots (17) are also associated with star-forming
  regions. We present comparisons to the thermal $^{13}$CO(1--0)
  emission as well as to other surveys of class II CH$_3$OH and H$_2$O
  maser emission. The catalogue attempts to present associations to
  astrophysical sources where available, and the full catalogue is
  available in electronic form.}
{This OH maser catalogue presents a unique resource of stellar and
  interstellar masers in the northern hemisphere. It provides the
  basis for a diverse range of follow-up studies from envelopes around
  evolved stars to star-forming regions and Supernova remnants.}
\keywords{Stars: formation -- ISM: clouds -- ISM: kinematics and
  dynamic -- stars: evolution -- catalogs -- Masers} \titlerunning{OH
  maser in THOR}

\maketitle

\section{Introduction}
\label{intro}

Studying the Milky Way in various tracers over large spatial scales
allows us to investigate the structure and physical processes in our
home Galaxy in great depth. The previous decades have seen a multitude
of Galactic plane surveys from long centimeter wavelengths (e.g.,
\citealt{becker1995,stil2006,hoare2012}) via surveys at (sub)mm
wavelengths (e.g.,
\citealt{dame2001,jackson2006,schuller2009,schuller2017,rigby2016})
and far-/mid-/near-infrared wavelengths (e.g.,
\citealt{egan2003,skrutskie2006,churchwell2009,carey2009,molinari2016b})
to high-energy gamma ray surveys (e.g., \citealt{hess2018}) with
almost all accessible continuum bands and many spectral lines
covered. Most previous surveys at mm to (sub)mm wavelengths sample
largely thermal emission from gas and dust. However, recently more and
more surveys of the non-thermal maser emission of our Milky Way were
conducted. For example, CH$_3$OH class I masers were studied by the
Millimeter Astronomer's Legacy Team-45\,GHz survey, MALT-45
\citep{jordan2013,jordan2015,jordan2017}, water masers by the H$_2$O
southern Galactic plane survey (HOPS)
\citep{walsh2011,walsh2012,walsh2014,longmore2017} or CH$_3$OH class
II masers by the Methanol Multi-Beam (MMB) survey
\citep{green2009,green2017,breen2015,breen2018}.

One other important maser species are the OH masers found in this
molecule's ground state transitions at $\sim$18\,cm wavelength in the
radio band (e.g., \citealt{weaver1965,elitzur1992b}). OH masers have
been surveyed in the past decades in various ways, mostly via pointed
follow-up observations toward IRAS sources (a summary of most of the
surveys can be found in \citet{mu2010} and
\citealt{qiao2014}). Unbiased surveys of the most abundant of these OH
masers, the 1612\,MHz transition, have been conducted in the southern
and northern hemisphere with the Australia Telescope Compact Array
(ATCA) and the Very Large Array (VLA) by
\citet{sevenster1997a,sevenster1997b,sevenster2001}. Furthermore,
\citet{dawson2014} and \citet{qiao2016} used the Parkes single-dish
telescope to survey all four OH maser transitions in the southern
hemisphere. While the VLA and ATCA studies only covered a single maser
transition at moderate sensitivity, the Parkes survey covers only
parts of the southern hemisphere. Therefore, covering the northern
hemisphere in all four OH maser transitions at good sensitivity and
spatial resolution is an important step to get an unbiased picture of
the OH maser distribution and the associated physical processes. Here,
we present the results of the OH maser detections within ``The
HI/OH/Recombination line survey of the inner Milky Way (THOR)''
conducted with the VLA. The OH maser results of the pilot region
around W43 have been presented in \citet{walsh2016}. An overview of the
entire survey can be found in \citet{beuther2016}.

A survey covering all four OH transitions is pertinent for
interstellar medium (ISM), star formation and evolved star studies
because they trace very different physical processes. While the most
abundant 1612\,MHz OH maser transition is typically known to favorably
trace the expanding shells of evolved stars (e.g.,
\citealt{wilson1968,hyland1972,elitzur1976b,elitzur1992b}), the two
main line masers at 1665 and 1667\,MHz are more often found toward
star-forming regions (e.g.,
\citealt{reid1981,argon2000,fish2003,qiao2014,qiao2016}), but can also
be found toward OH/IR stars (e.g., \citealt{lewis1997}). The least
frequent 1720\,MHz maser is typically excited by shocks and found
toward either star-forming regions or supernovae remnants (e.g.,
\citealt{gray1992,pavlakis1996b,lockett1999,caswell1999,caswell2004,claussen1999,wardle2002}). However,
these maser associations are not exclusive, and there are also several
cases where different OH maser lines exist toward the same source
(e.g., \citealt{caswell2013,walsh2016}). Furthermore, detecting a
maser line does not allow a priori to unambiguously indicate the
nature of the region it arises from, and in most cases complementary
data are needed to elucidate this issue. \citet{mu2010} compiled from
the literature 3249 OH maser sources, and they attributed 7.4, 78.9
and 13.7\% to interstellar, stellar or unknown origin category,
respectively.  In summary, studying OH masers is a unique tool to
sample important evolutionary stages of stellar evolution, i.e.,
star formation, evolved stars and supernova remnants.

\section{The THOR survey and its data}
\label{data}

The THOR survey covers Galactic longitudes from 14\pdeg 35 to 67\pdeg
25 within Galactic latitudes of $\pm$1\pdeg 25 with the VLA in the
C-array configuration at the radio L-band between 1 and 2\,GHz
frequencies. With the flexible WIDAR correlator, we simultaneously
observe the entire band in the continuum emission as well as the
spectral lines of HI, the four OH transitions as well as several
hydrogen recombination lines (e.g.,
\citealt{bihr2015b,walsh2016,wang2018,rugel2018,rugel2019}). A
description of the survey, its design and specification is given in
\citet{beuther2016}.  A link to the first data release and additional
publications about various aspects of the survey can be found on the
project website\footnote{http://www.mpia.de/thor}.

The OH data were observed in right-hand-circular and
left-hand-circular polarization, but not in any of the
cross-products. We produced the Stokes I total intensity maps of the
OH data, and we focus on the final data products of the four OH
transitions at 1612, 1665, 1667 and 1720\,MHz (see also
\citealt{walsh2016,rugel2018}). The whole survey was conducted in
three steps: a pilot study around the high-mass star-forming region
W43 \citep{bihr2015b,walsh2016}, then the first and second half of the
survey separately (see \citealt{beuther2016} for details). While the
pilot study as well as the second half of the survey cover all four OH
transitions, due to a problem in the setup, the first half of the
survey unfortunately did not cover the 1667\,MHz transition (longitude
ranges roughly $14\pdeg35<l<29\pdeg25$, $31\pdeg5<l<37\pdeg9$ and
$47\pdeg1<l<51\pdeg2$). For the data calibration and imaging strategy
and procedures, we refer to \citet{walsh2016}, \citet{beuther2016} and
\citet{rugel2018}. Continuum subtraction was conducted prior to
  the imaging.

Using a Briggs weighting scheme with a robust factor of 0, the
spatial resolution of the OH data cubes varies between roughly
$12.5''$ and $19''$ (see Table 2 in \citet{beuther2016}, and the
spectral resolution of the final data cubes is 1.5\,km\,s$^{-1}$. The
typical rms noise level of the data is $\sim$10\,mJy\,beam$^{-1}$, but
it can increase close to strong maser features because the
interferometric THOR data have a dynamic range limit of $\sim$900
  \citep{walsh2016}, and peak flux densities greater 10\,Jy are found
  toward several sources (Table \ref{catalogue}).

\section{Results}
\label{results}

The main product of this paper is the catalogue of OH masers in
the THOR survey. We follow the nomenclature given by \citet{walsh2016}
for the THOR pilot region W43. The term ``maser spot'' refers to a
single peak in the spectrum which typically originates from parcels of
gas with sizes around $10^{14}$\,cm or $\sim$10\,AU (e.g.,
\citealt{elitzur1992b}). Such maser spots may be grouped in regions
with sizes around $10^{16-17}$\,cm ($10^3-10^4$\,AU) which we call
``maser sites''. Given our spatial resolution and typical distances,
maser spots and maser sites are typically unresolved in our
observations. Our definition of maser spot is called in some other
works also ``maser feature'' or ``maser component''.

\begin{figure*}[htb]
  \includegraphics[angle=-90,width=0.99\textwidth,viewport=220 100 350 700]{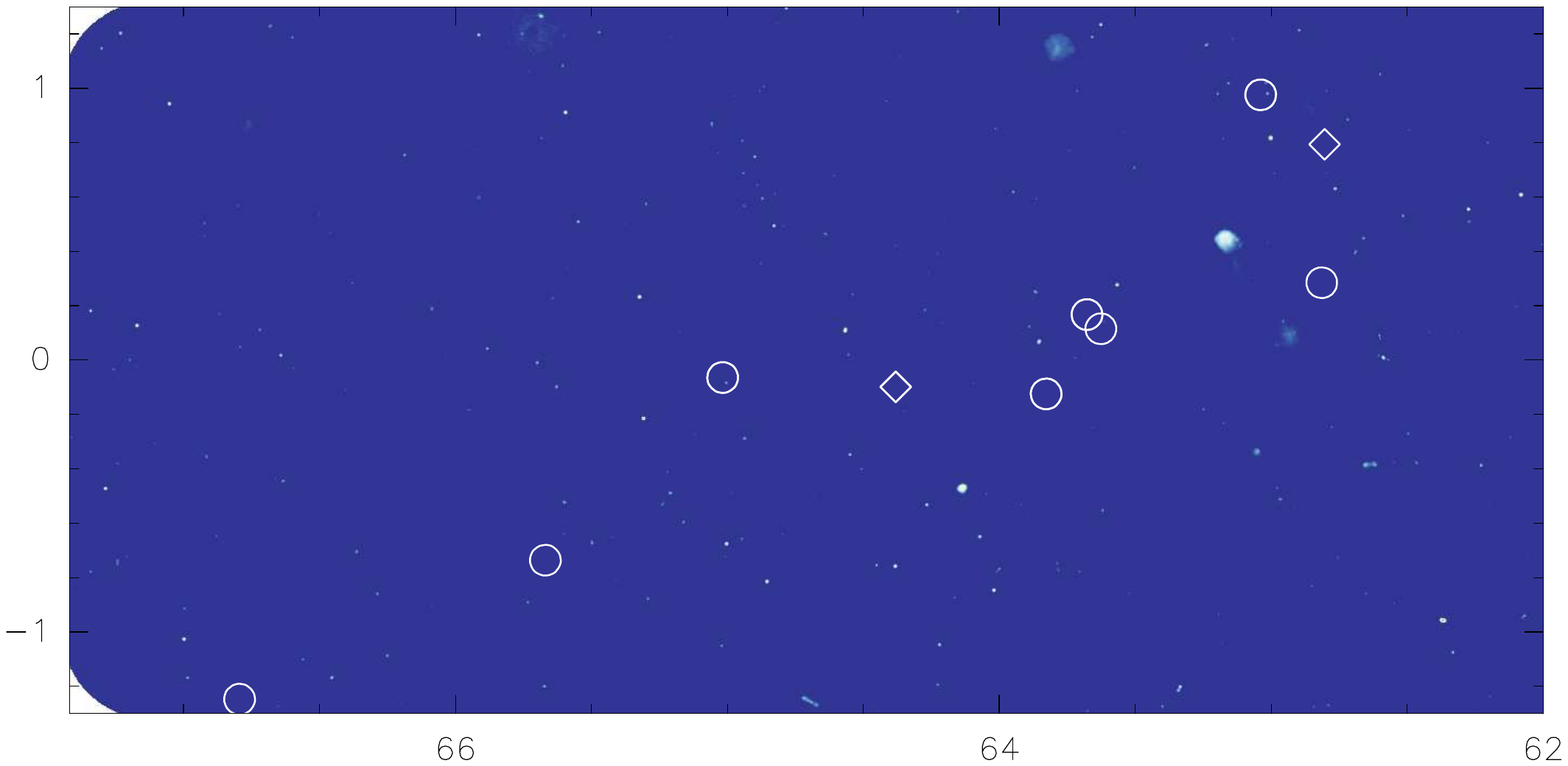}
  \includegraphics[angle=-90,width=0.99\textwidth,viewport=80 100 350 700]{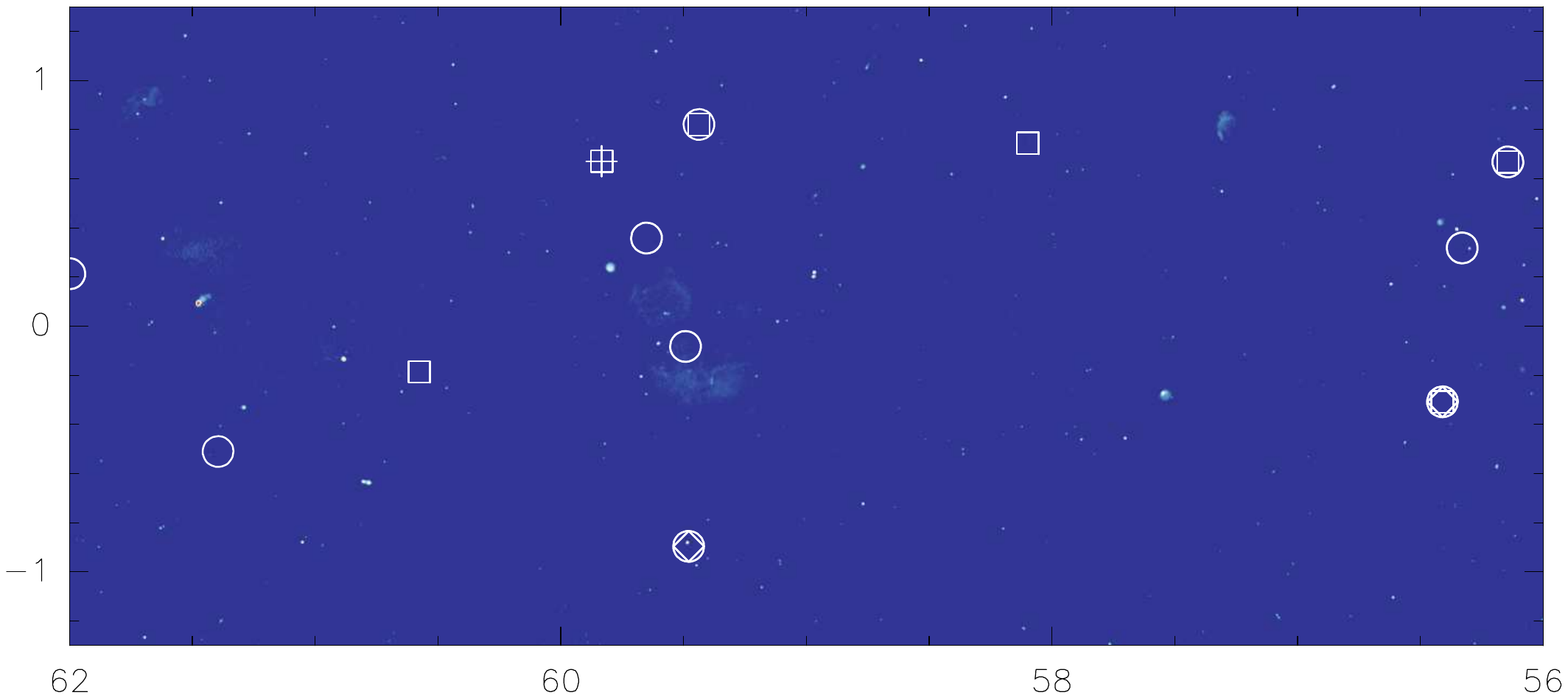}
  \includegraphics[angle=-90,width=0.99\textwidth,viewport=80 100 503 700]{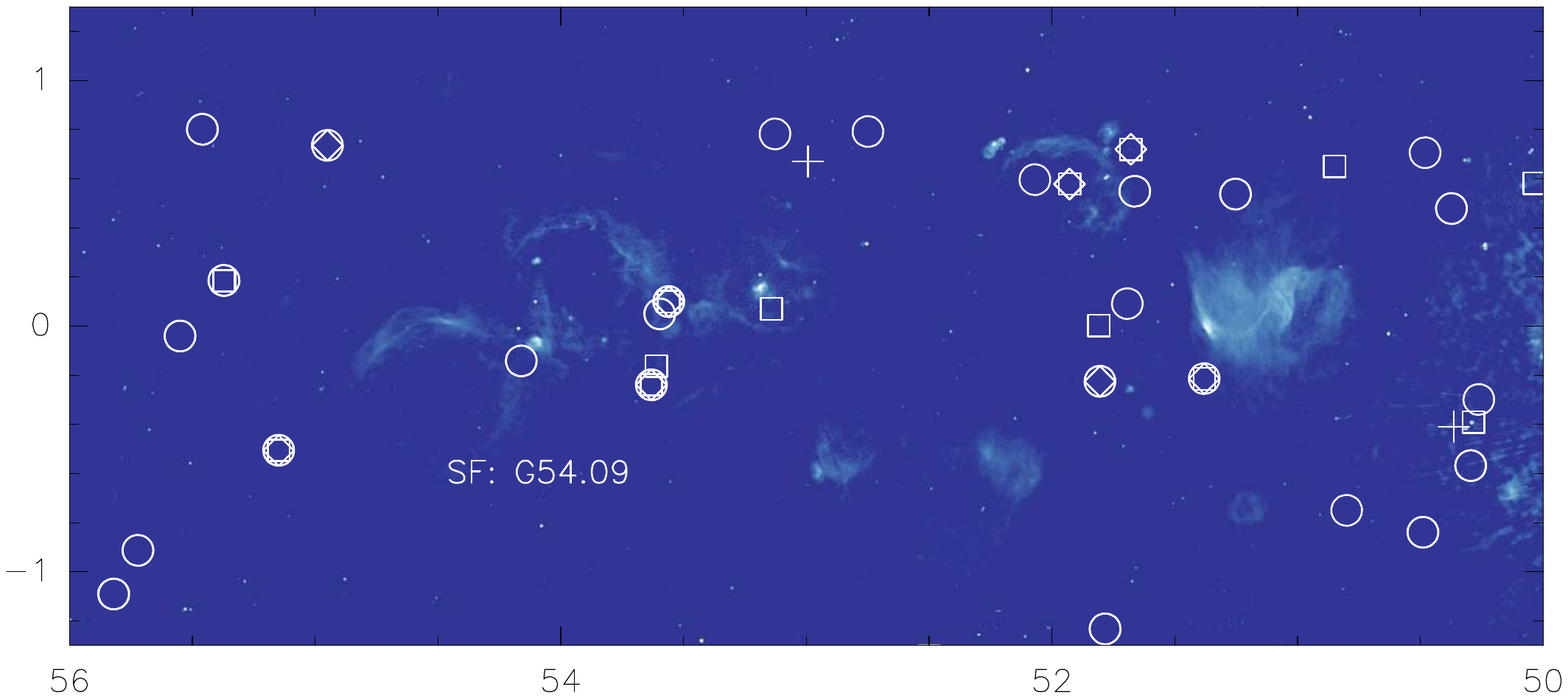}
  \caption{THOR OH maser spot positions overlaid on the 1.4\,GHz
    continuum map in Galactic coordinates. The circles, squares,
    diamonds and plus-signs mark the 1612, 1665, 1667, and 1720\,MHz
    masers, respectively. A few example regions are labeled.}
\label{samplefig1} 
\end{figure*} 

\setcounter{figure}{0}
\begin{figure*}[htb]
  \includegraphics[angle=-90,width=0.99\textwidth,viewport=250 100 350 700]{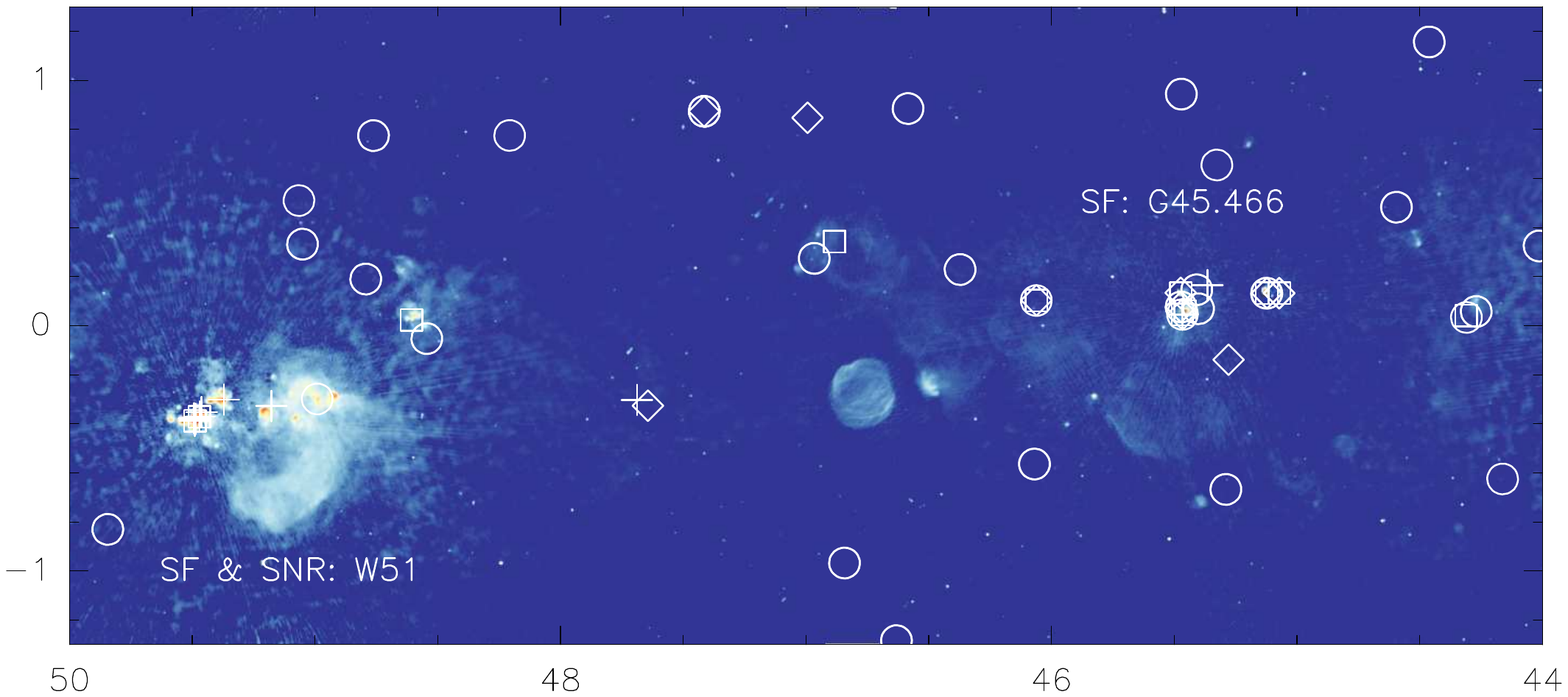}
  \includegraphics[angle=-90,width=0.99\textwidth,viewport=80 100 350 700]{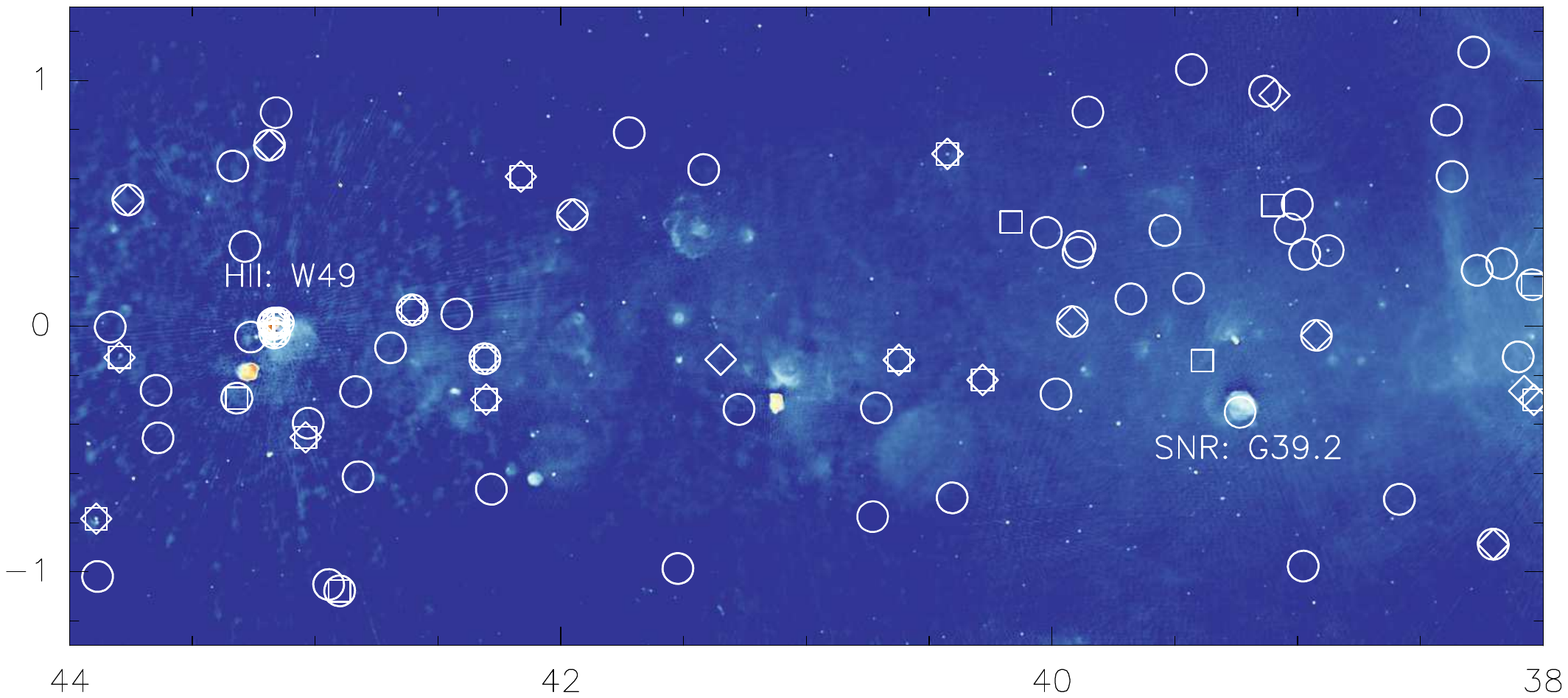}
  \includegraphics[angle=-90,width=0.99\textwidth,viewport=80 100 507 700]{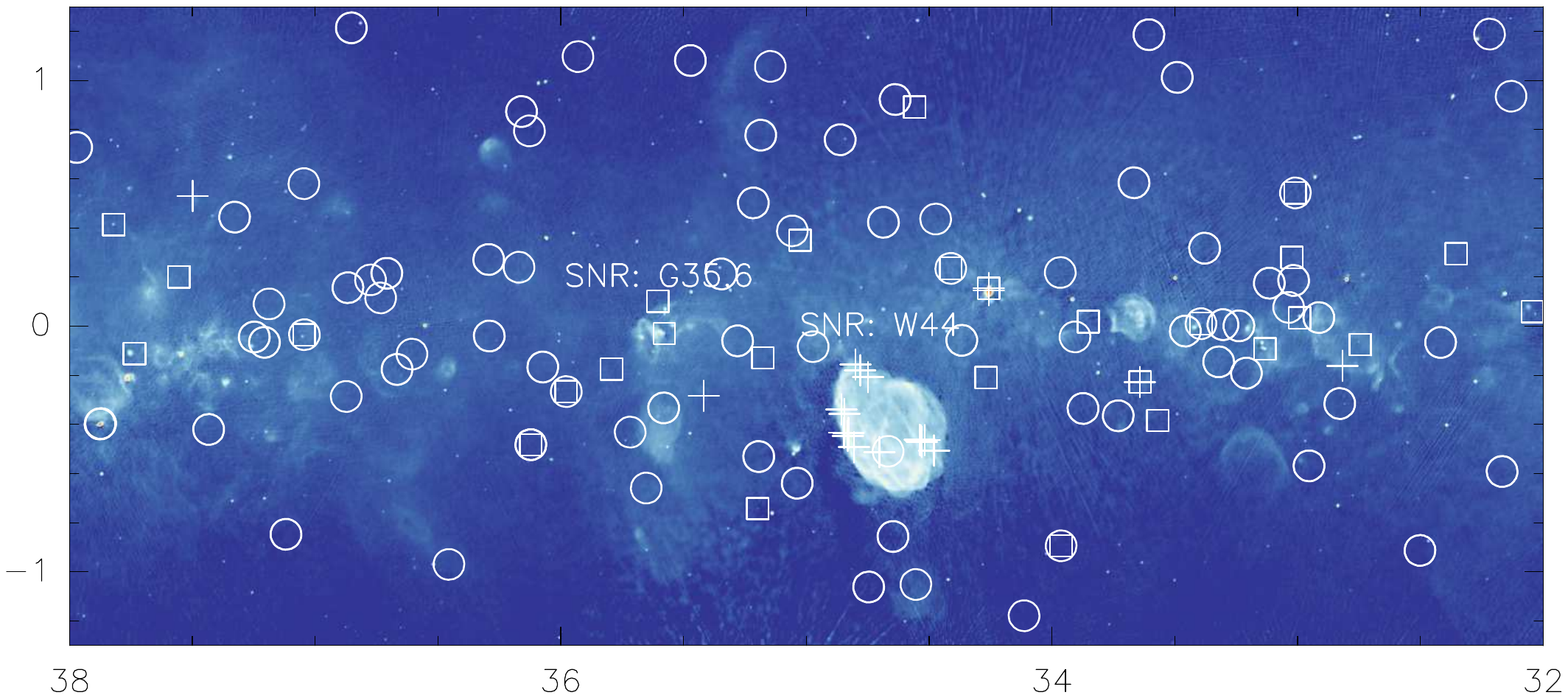}
  \caption{Fig.~\ref{samplefig1} continued.}
\end{figure*} 

\setcounter{figure}{0}
\begin{figure*}[htb]
  \includegraphics[angle=-90,width=0.99\textwidth,viewport=250 100 350 700]{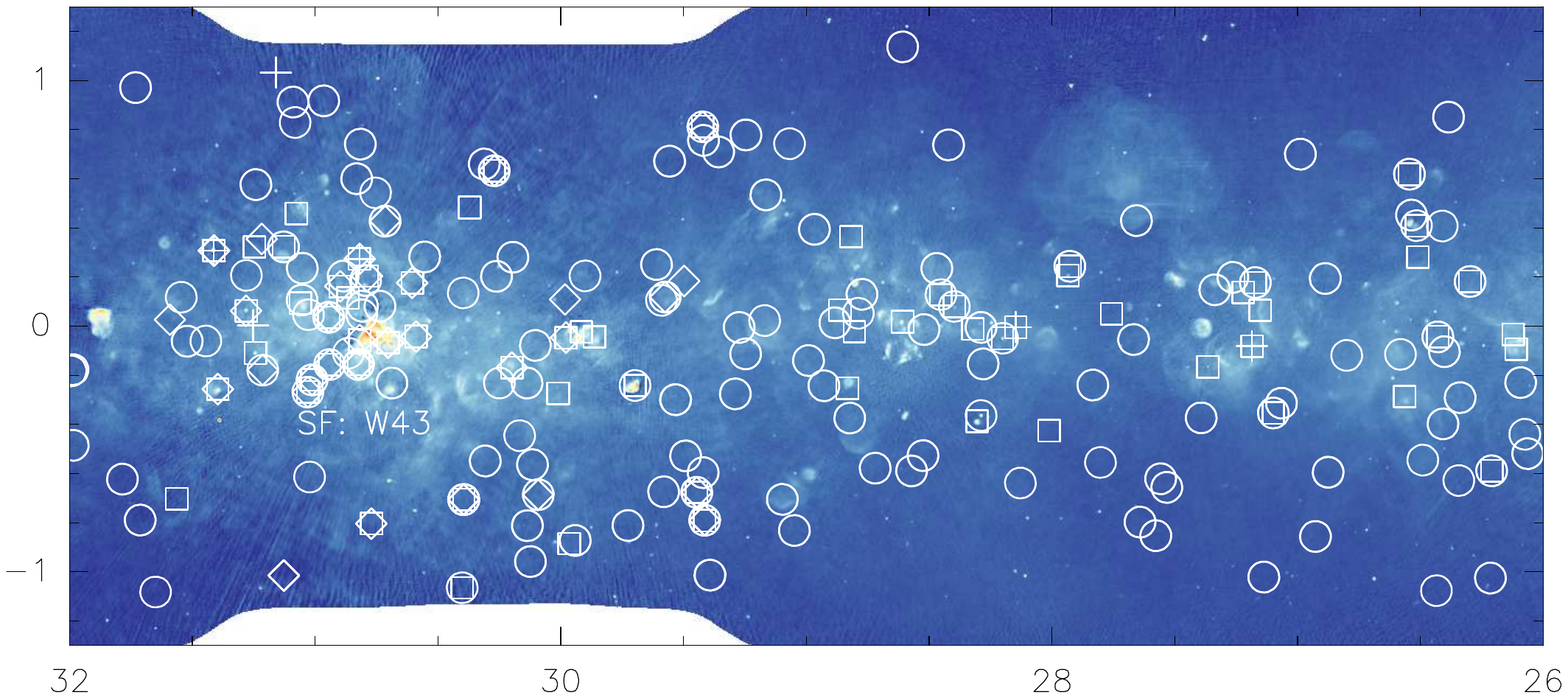}
  \includegraphics[angle=-90,width=0.99\textwidth,viewport=80 100 350 700]{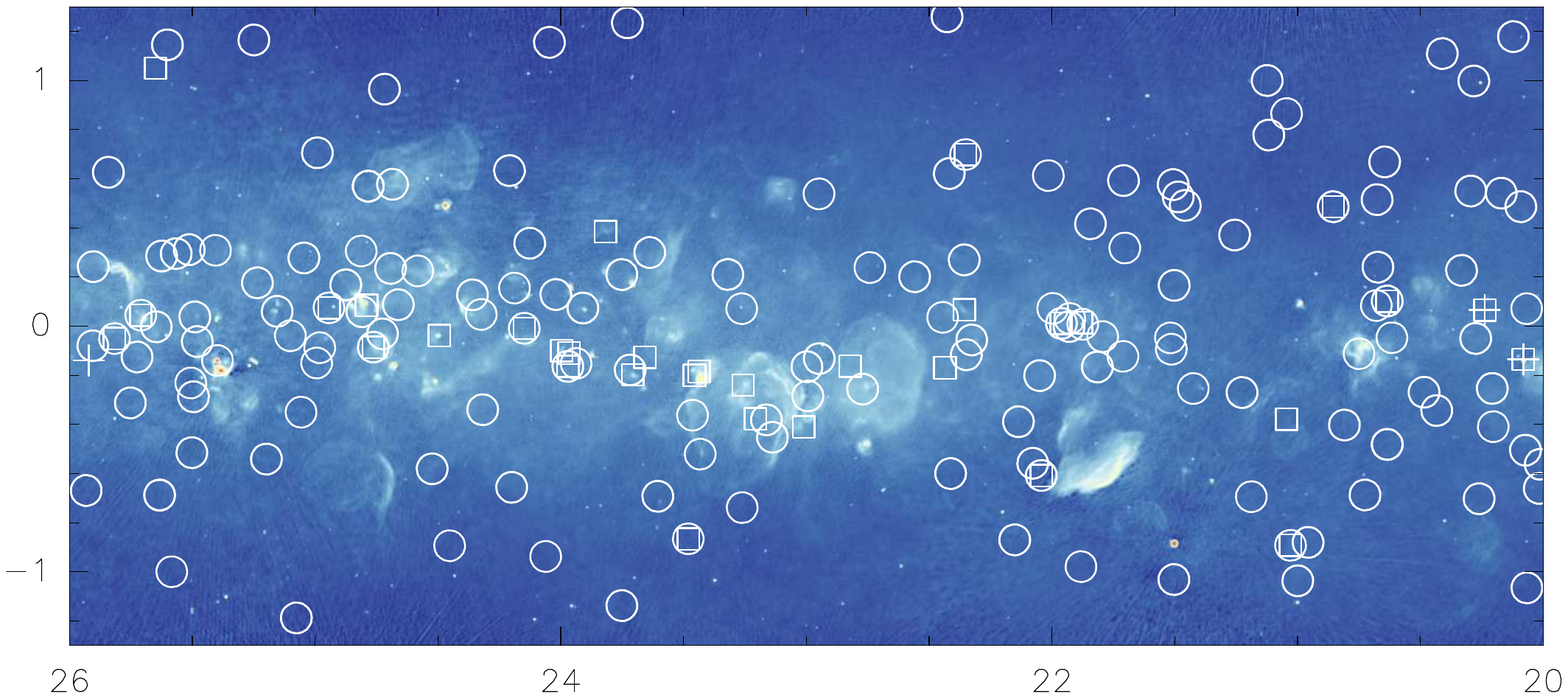}
  \includegraphics[angle=-90,width=0.99\textwidth,viewport=60 100 505 700]{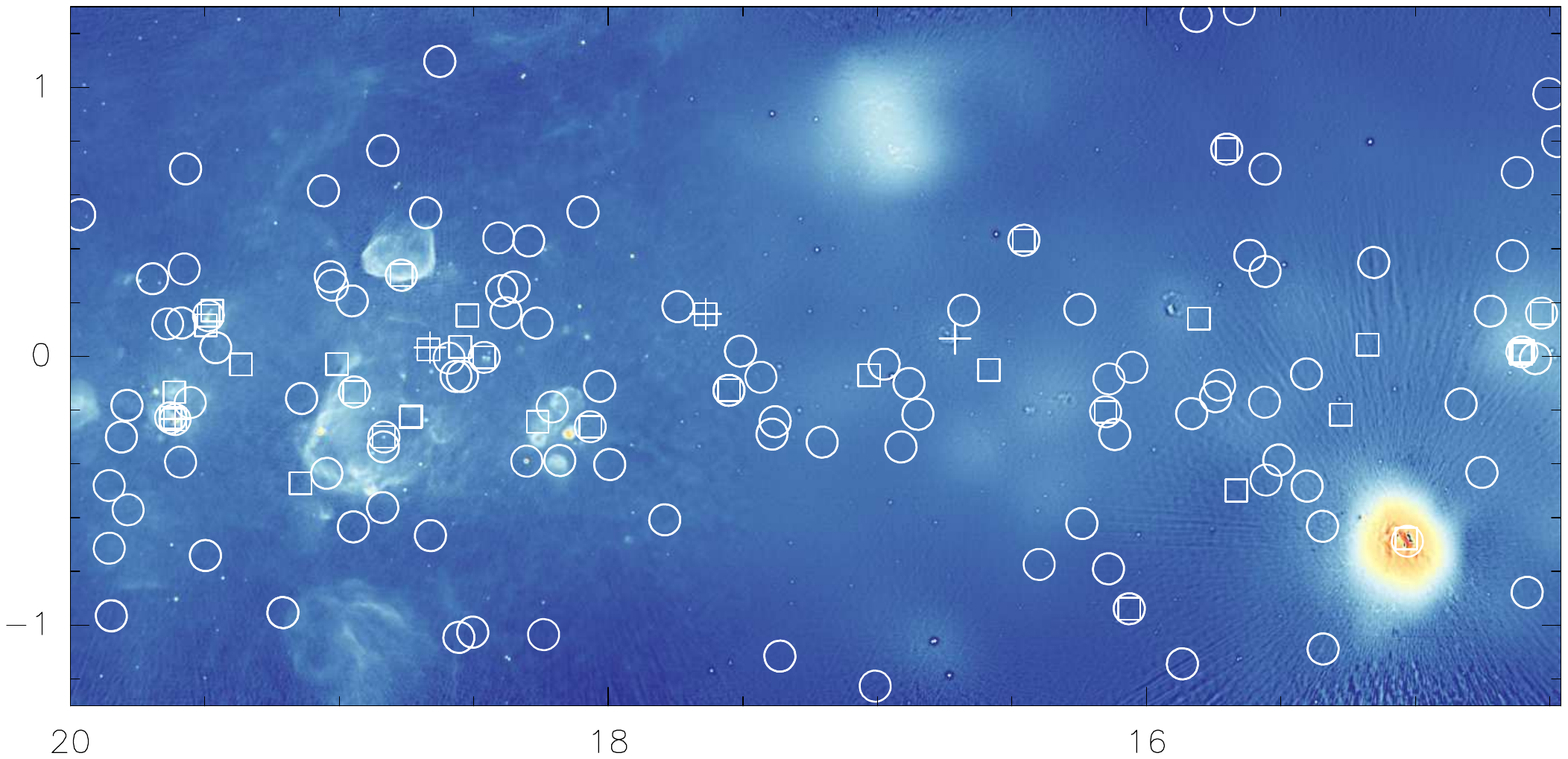}
  \caption{Fig.~\ref{samplefig1} continued.}
\end{figure*} 

\begin{figure*}[ht]
\includegraphics[width=0.24\textwidth]{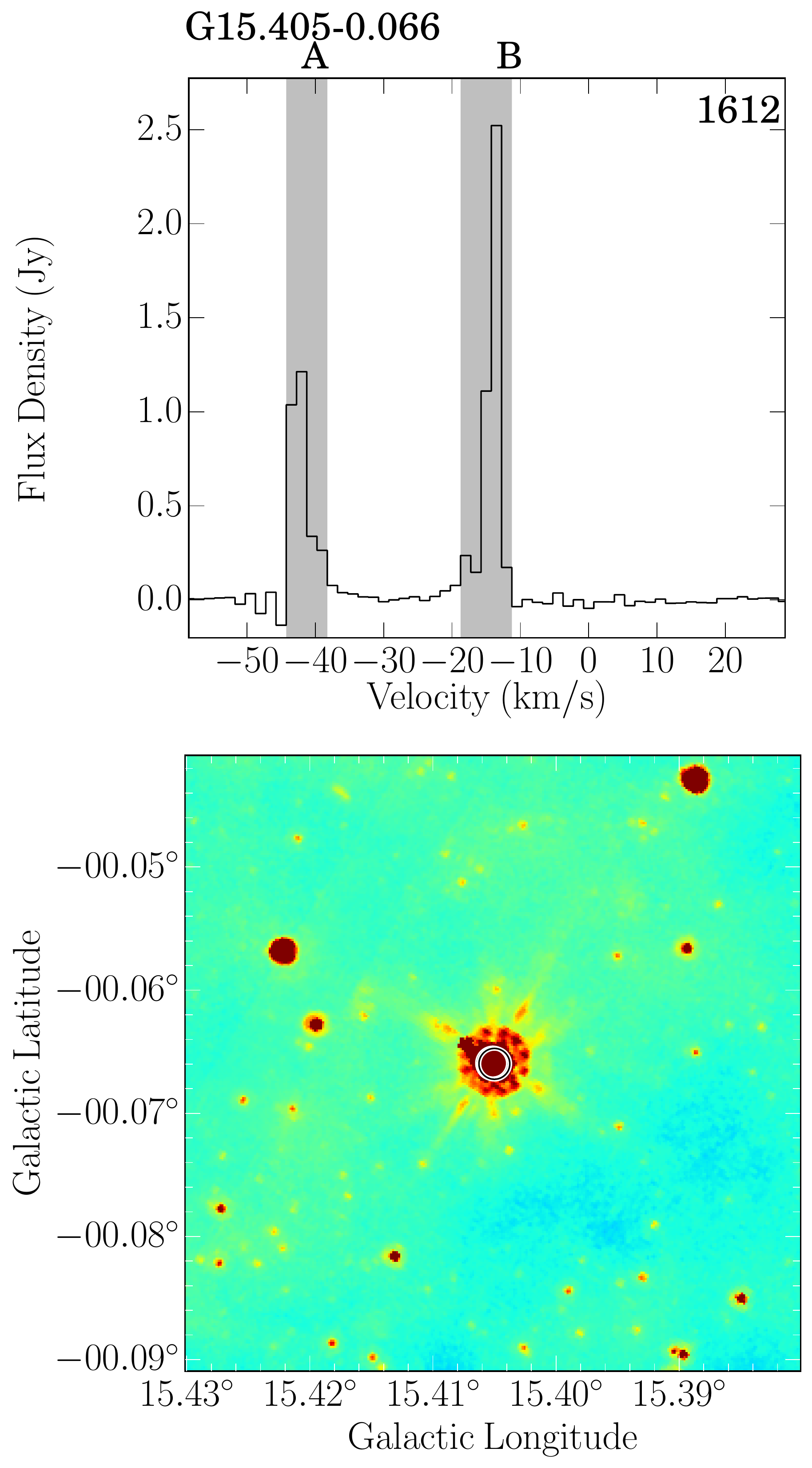}
\includegraphics[width=0.25\textwidth]{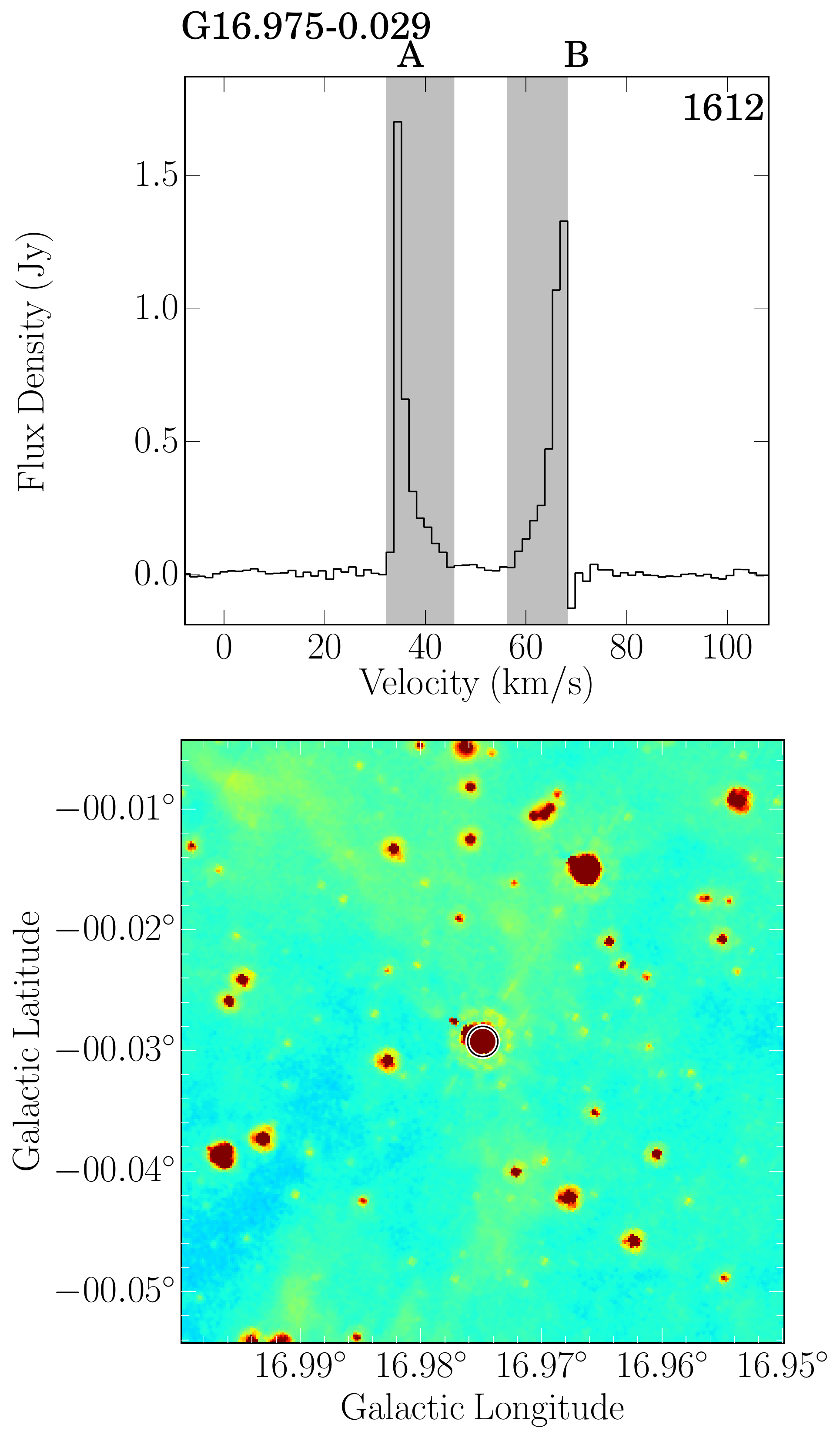}
\includegraphics[width=0.245\textwidth]{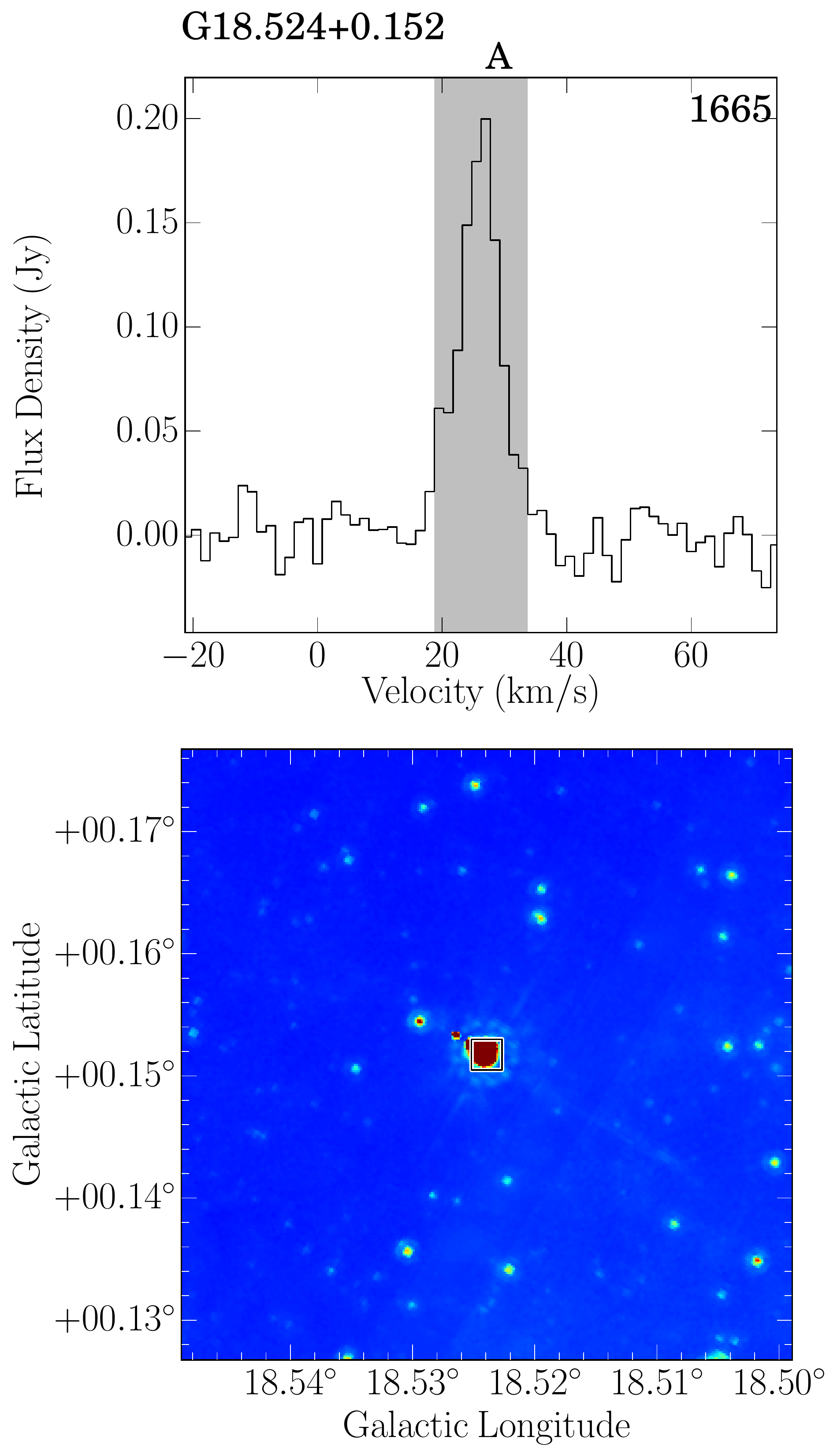}
\includegraphics[width=0.24\textwidth]{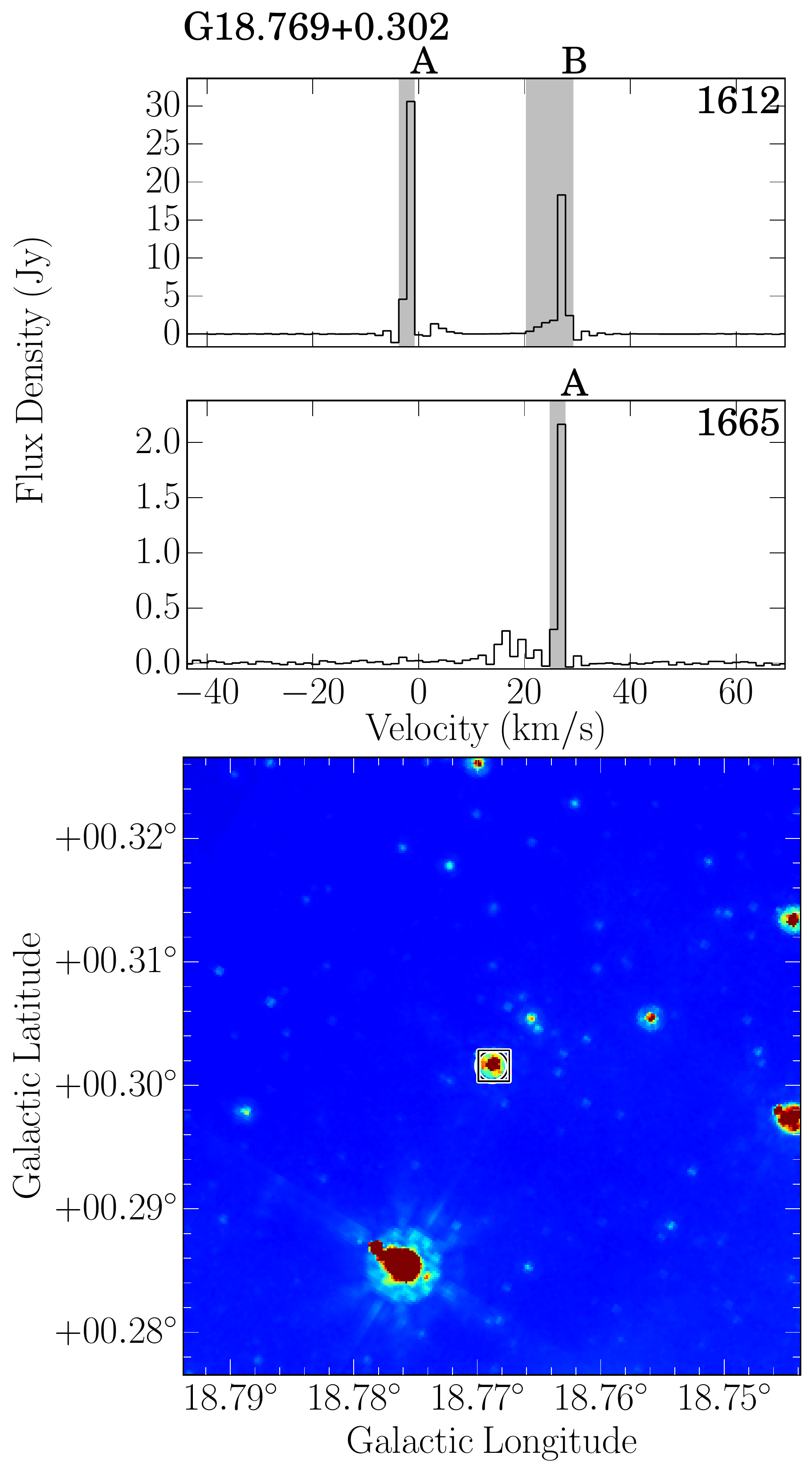}
\includegraphics[width=0.24\textwidth]{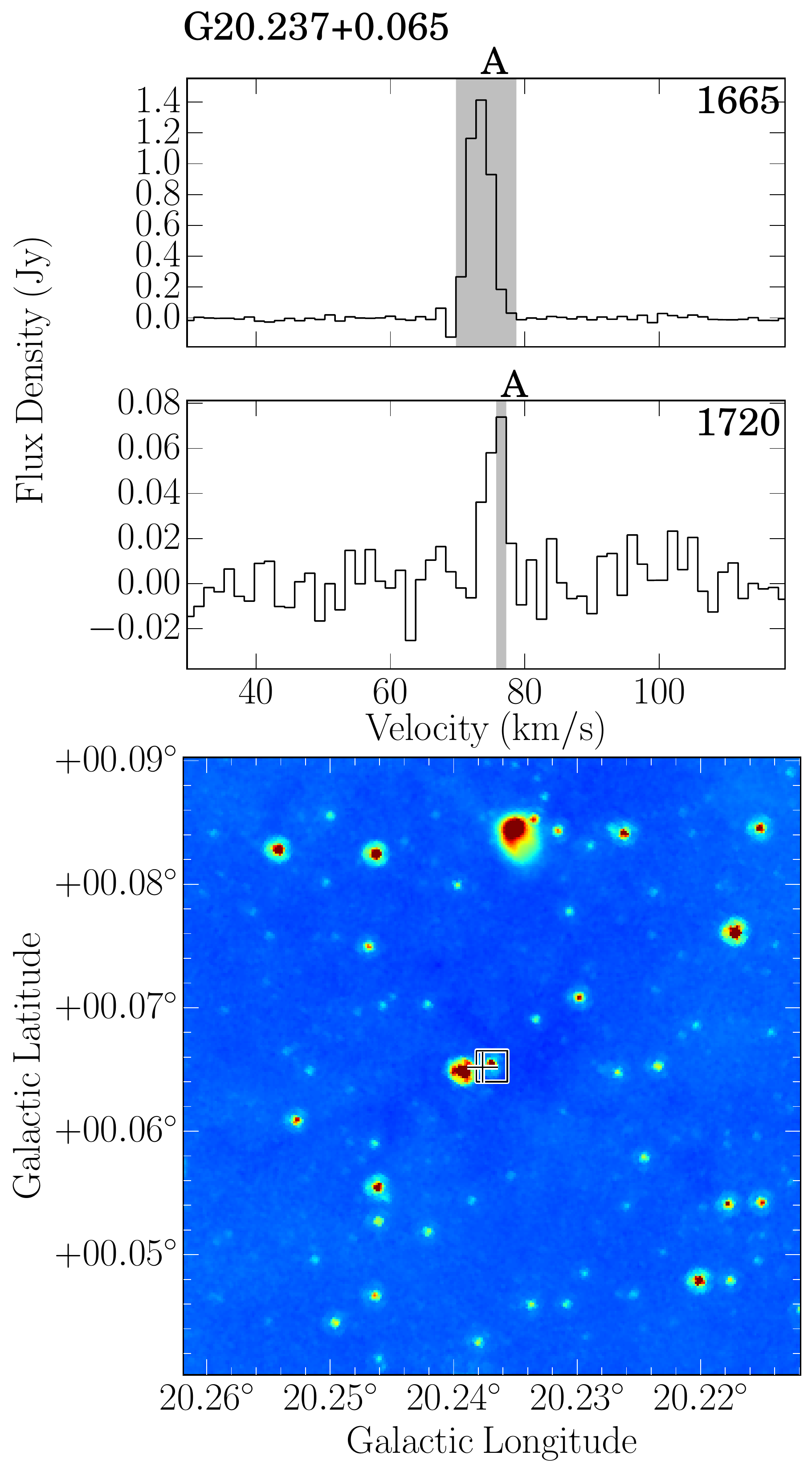}
\includegraphics[width=0.24\textwidth]{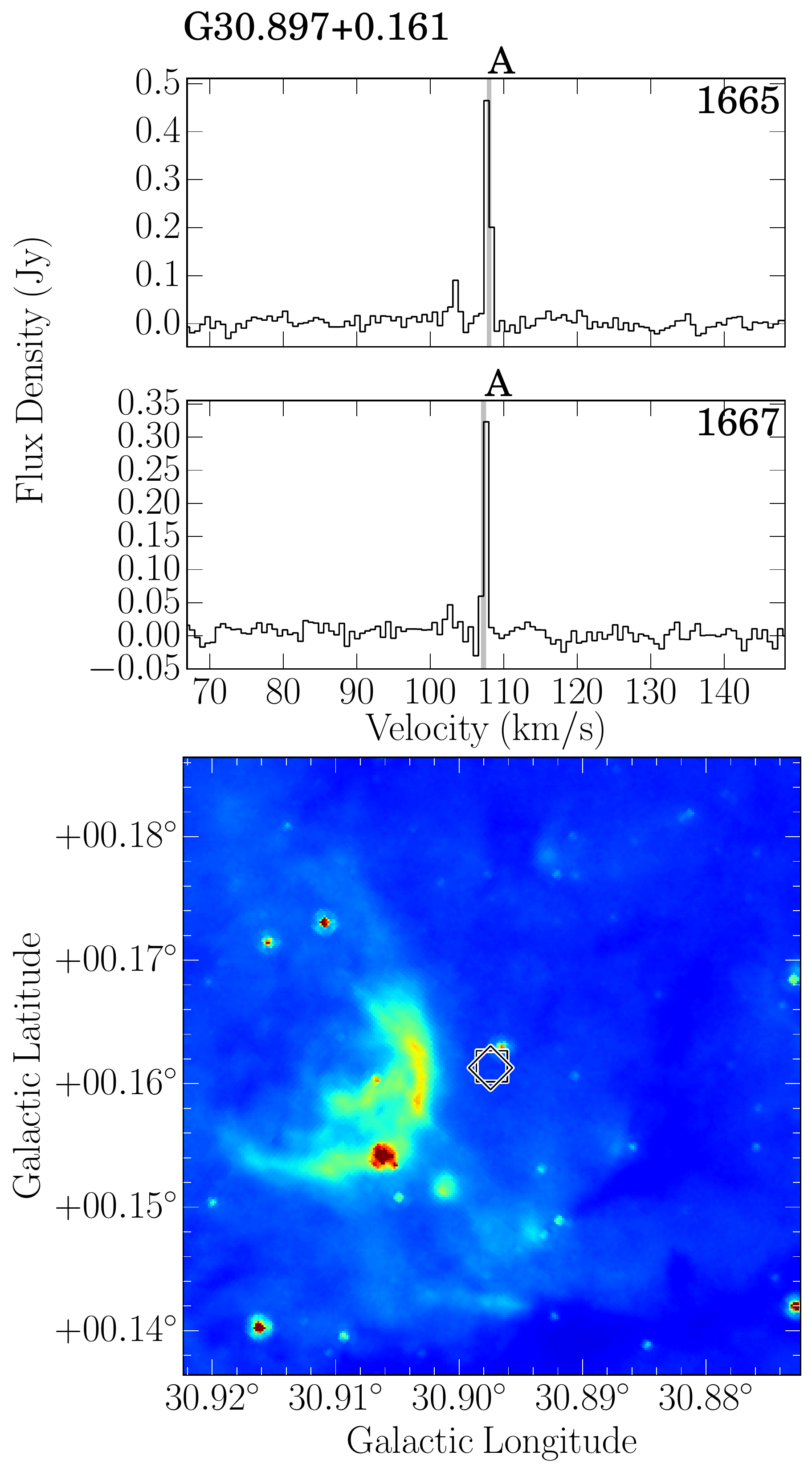}
\includegraphics[width=0.24\textwidth]{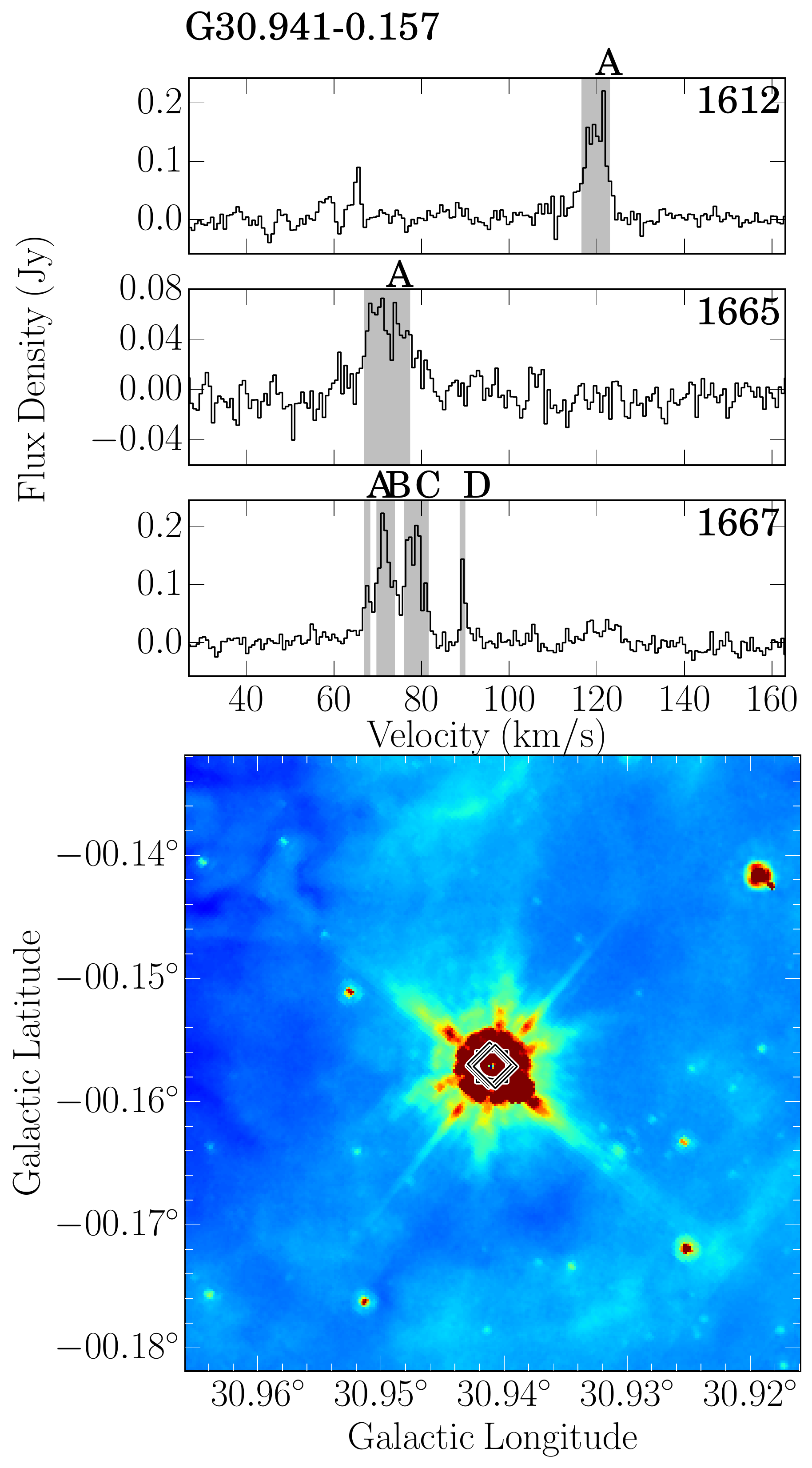}
\includegraphics[width=0.24\textwidth]{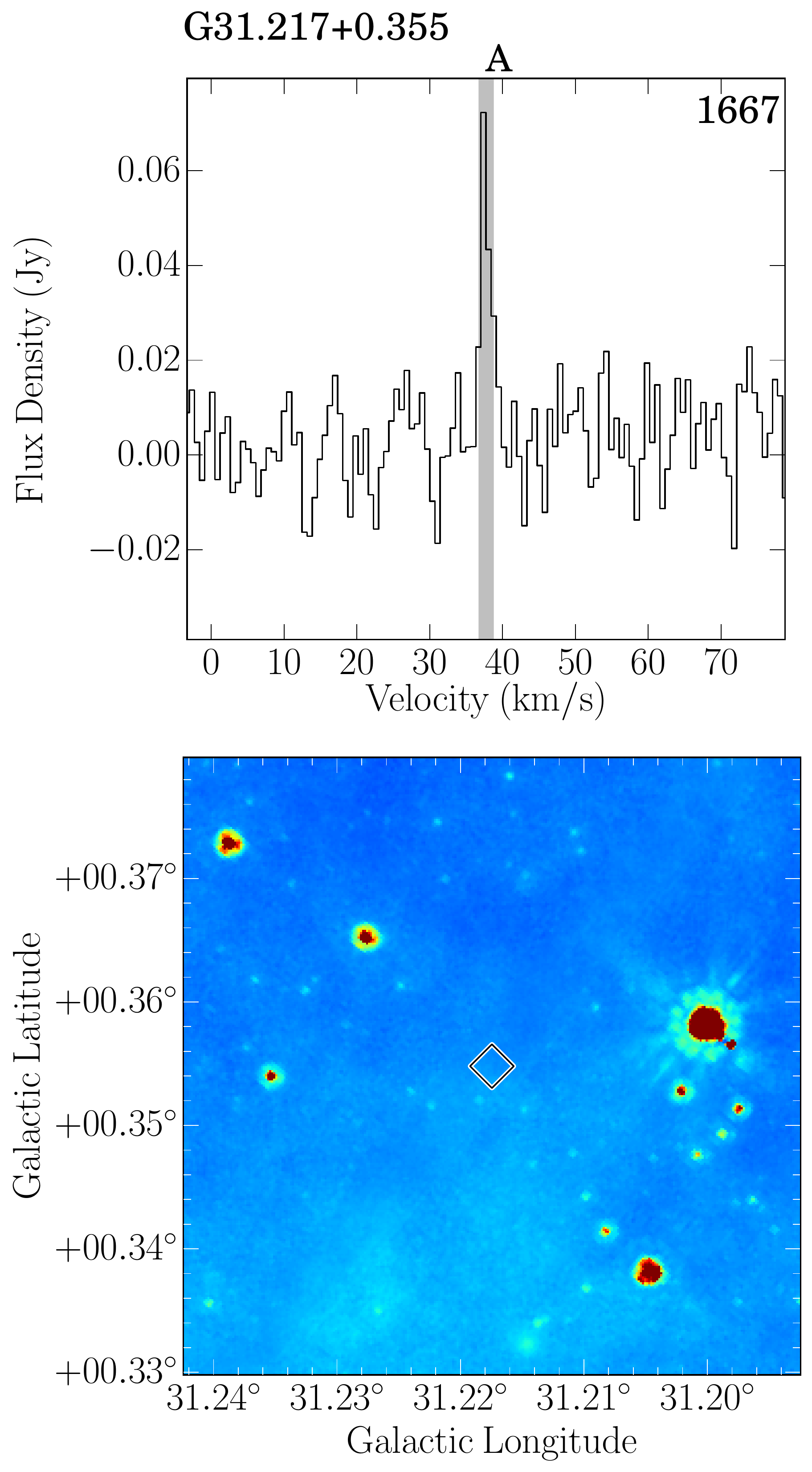}
\includegraphics[width=0.24\textwidth]{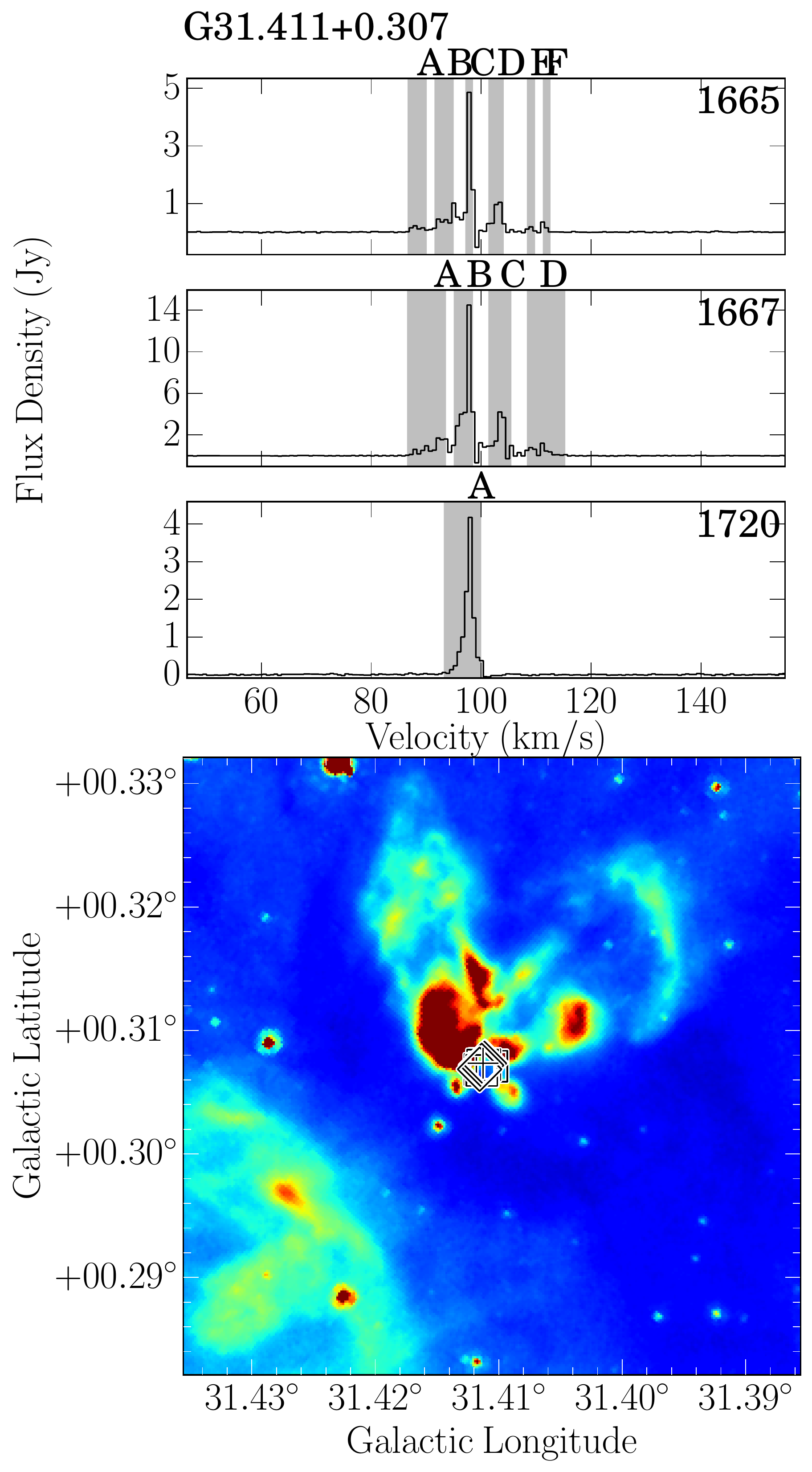}
\includegraphics[width=0.25\textwidth]{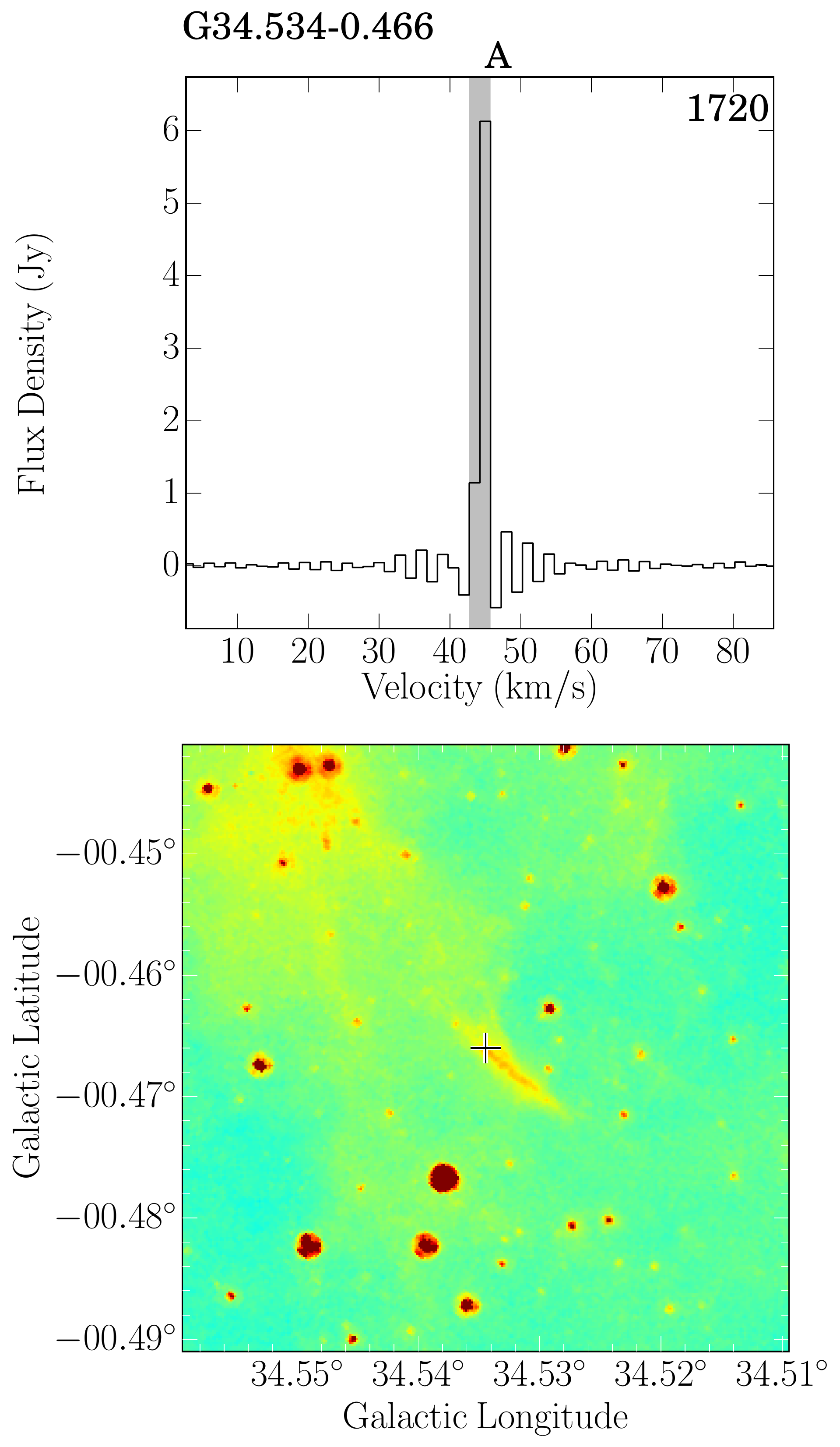}
\includegraphics[width=0.24\textwidth]{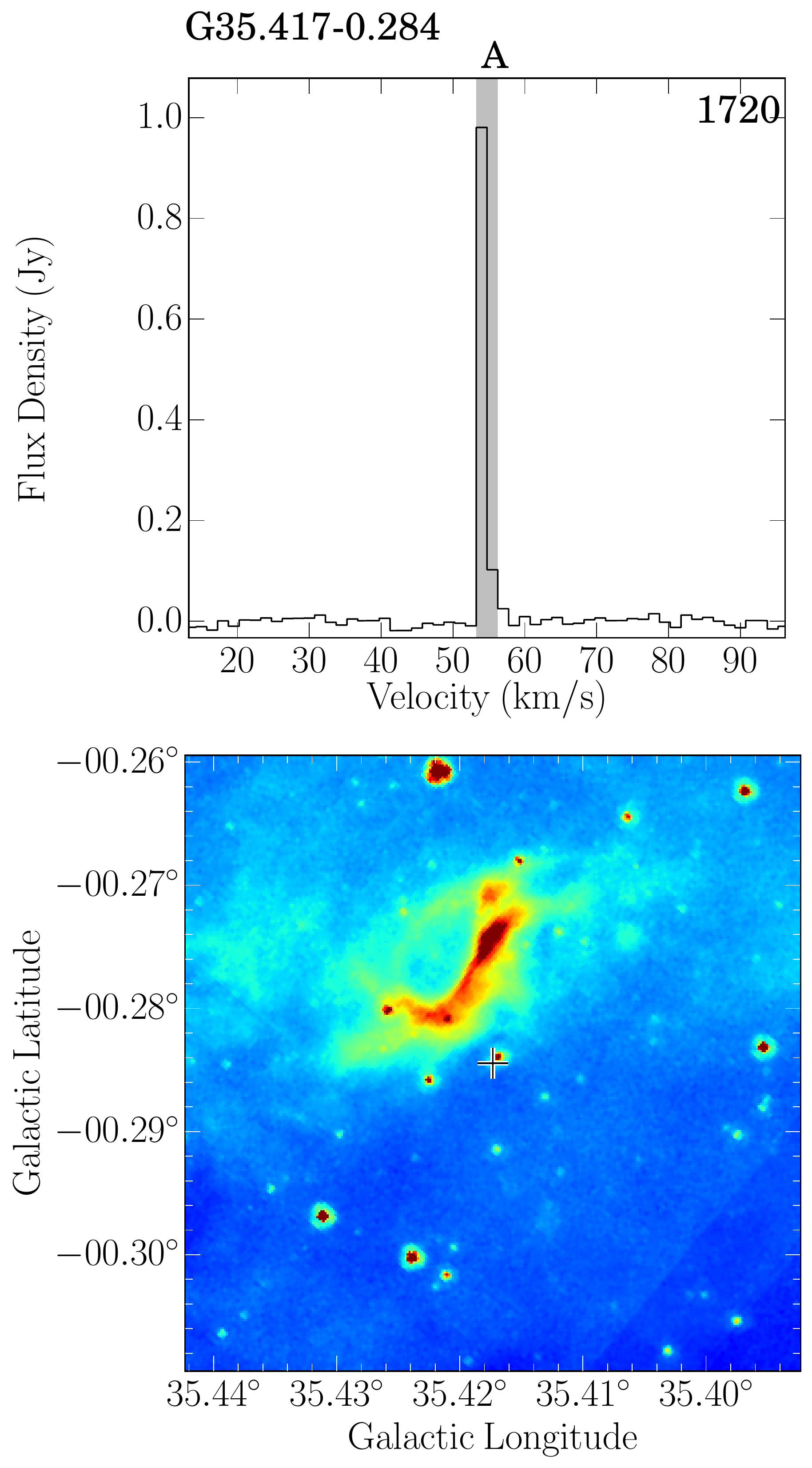}
\includegraphics[width=0.24\textwidth]{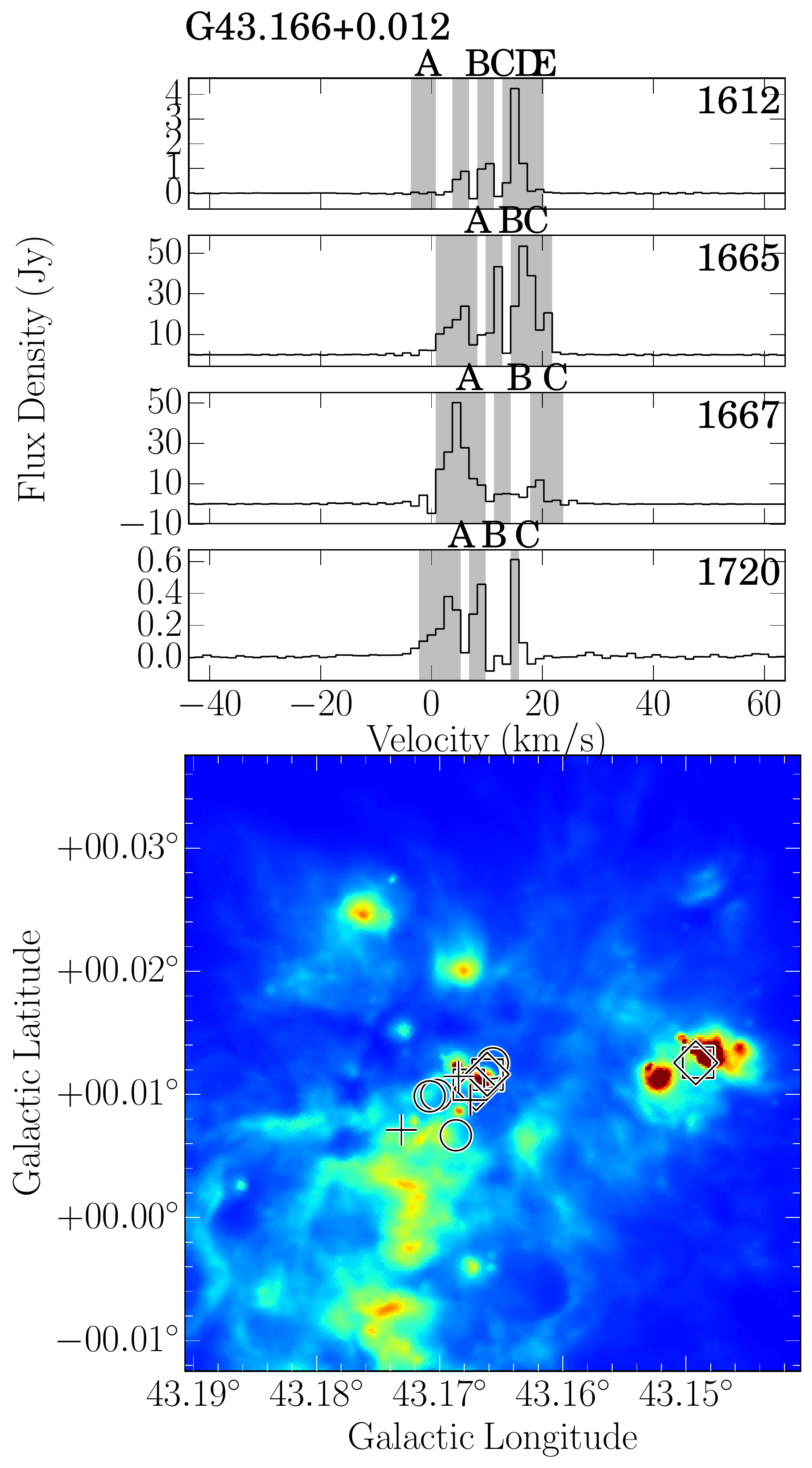}
\caption{Examples of individual OH spectra and 8\,$\mu$m images for
  different OH maser lines (remaining maser sites are presented in the
  appendix). For each maser site, the top panel(s) show the
  spectrum/spectra identified in this region, and the bottom panels
  present the corresponding GLIMPSE 8\,$\mu$m image
  \citep{benjamin2003,churchwell2009}. Where GLIMPSE data are not
  available, the corresponding WISE data \citep{wright2010} are
  used. The grey-shaded parts in the spectra outline the velocity
  regimes of the individual maser spots, and the circles, squares,
  diamonds, and plus-signs mark the positions of the 1612, 1665, 1667,
  and 1720\,MHz maser spots, respectively (spatially overlapping ones
  are sometimes difficult to visually separate).}
\label{example}
\end{figure*}

\clearpage

\begin{table*}[htb]
\caption{THOR maser spot catalogue (example). The full Table is available at CDS.}
\begin{tabular}{lccccccccccc}
\hline
\hline
Name                  & R.A.            & Dec.            &$S_{\rm peak}$& rms  & $S_{\rm int}$ & $\varv_{\rm peak}$ & $\varv_{\rm min}$   & $\varv_{\rm max}$ & $\Delta x$ & $\Delta y$ & ass.\\
                      &J2000.0        &  J2000.0      &  Jy       &mJy &$\rm{Jy}\frac{\rm{km}}{\rm{s}}$ &$\frac{\rm{km}}{\rm{s}}$ &$\frac{\rm{km}}{\rm{s}}$&$\frac{\rm{km}}{\rm{s}}$ &$''$ & $''$ & \\ 
\hline
  G14.389-0.020-1665A &   18:16:43.732  &  -16:27:03.33  &  0.073 &  3   &   0.111 &  24.0    &    23.25    &  24.75    &  0.3  &  0.3  &SF-Sim,MMB,rc\\
                                                                                                                                        & \\
  G14.431-0.033-1612A &   18:16:51.510  &  -16:25:12.01  &  0.366 &  8   &   1.977 &  33.0    &    27.75    &  35.25    &  0.2  &  0.2  & ES-D,BS\\  
  G14.431-0.033-1612B &   18:16:51.518  &  -16:25:11.63  &  0.660 &  8   &   2.042 &  39.0    &    38.25    &  44.25    &  0.1  &  0.1  & \\  
                                                                                                                                        & \\
  G14.455+0.014-1612A &   18:16:44.091  &  -16:22:36.96  &  0.104 &  5   &   0.139 &  82.5    &    80.25    &  83.25    &  0.2  &  0.2  & ES-D,BS\\  
  G14.455+0.014-1612B &   18:16:44.077  &  -16:22:37.79  &  0.067 &  5   &   0.099 &  115.5   &    114.75   &  116.25   &  0.5  &  0.4  & \\  
                                                                                                                                        & \\
  G14.471+0.799-1612A &   18:13:53.604  &  -15:59:15.76  &  0.055 &  4   &   0.175 &  54.0    &    51.75    &  56.25    &  0.3  &  0.3  & Star-Sim\\  
                                                                                                                                        & \\
  G14.504+0.976-1612A &   18:13:18.550  &  -15:52:27.07  &  0.803 &  6   &   2.691 &  19.5    &    17.25    &  26.25    &  0.1  &  0.0  & U\\  
                                                                                                                                        & \\
  G14.530+0.160-1612A &   18:16:20.858  &  -16:14:24.47  &  0.390 &  5   &   0.965 &  84.0    &    81.75    &  84.75    &  0.1  &  0.1  & Star-Sim\\  
  G14.530+0.160-1665A &   18:16:20.835  &  -16:14:25.50  &  0.080 &  4   &   0.210 &  73.5    &    72.75    &  75.75    &  0.3  &  0.2  & \\  
  ...\\
\hline
\hline
\end{tabular}
{\footnotesize Notes: Given are the name (including the maser frequency and spot label A.B, etc.), Right Ascension and Declination, peak flux density rms and integrated flux density, peak velocity and minimum/maximum velocity of maser spot, and the positional uncertainties. Empty lines separate maser sites. \\ Associations (ass.) ES: evolved star, D: double peaked, D?: potentially double peaked with 2nd component either two weak to be firmly identified or outside the imaged velocity range, BS, RS or wkS: in the GLIMPSE images a bright, red or weak star is seen co-spatial with the maser site, SF: star formation, SNR: supernova remnant, PN: planetary nebula, Star: star, nPL: near a pulsar, Sim: according to Simbad database, rc: cm continuum emission from \citep{wang2018}. Previous literature entries are referenced as: Bel13 - \citealt{beltran2013}; Blo94 - \citet{blommaert1994}; Cas95 - \citet{caswell1995}; Cod10 - \citet{codella2010}; Dea07 - \citet{deacon2007}; Deg04 - \citet{deguchi2004}; DiF08 - \citet{difrancesco2008}; Fel02 - \citet{felli2002}; He05 - \citet{he2005}; Hil05 - \citet{hill2005}; Ima13 - \citet{imai2013}; Kwo97 - \citet{kwok1997}; Kur94 - \citet{kurtz1994}; Lou93 - \citet{loup1993}; Mot03 - \citet{motte2003}; MMB - Methanol Multibeam Survey \citet{caswell2010,green2010,breen2015}; Per09 - \citet{peretto2009}; Pes05 - \citet{pestalozzi2005}; Ros10 - \citet{rosolowsky2010}; Sev01 - \citet{sevenster2001}; Tho06 - \citet{thompson2006}; Urq09 - \citet{urquhart2009}; Wal98 - \citet{walsh1998}; Win75 - \citet{winnberg1975}.}
\label{catalogue}
\end{table*}

\subsection{Maser identification and completeness}

For the maser identification, we also follow the approach conducted
for the THOR pilot region in \citet{walsh2016}. As a starting point,
the package {\sc DUCHAMP} \citep{whiting2012} was used to identify
maser candidates. The data cubes have pixel sizes of $3''$ to sample
the beam linearly at least 4 times, mostly even 5-6 times (see
Sect.~\ref{data}). With the given rms and spectral resolution, {\sc
  DUCHAMP} searched for emission above twice the rms (typically
20\,mJy\,beam$^{-1}$) with additional criteria of at least 12 pixels
in a single channel, and at minimum 2 consecutive channels with
emission above the 2\,$\sigma$ level.

As outlined in \citet{walsh2016}, the {\sc DUCHAMP} maser
identification is furthermore limited by the varying noise with
respect to spatial position and spectral channel. Toward strong maser
sources, the channels including strong emission are not dominated by
thermal noise but rather by systematic errors caused by artifacts due
to the emission's side-lobes. To account for the noise variations, a
noise map was created, the emission map divided by the noise map, and
the {\sc DUCHAMP} algorithm was then run on the resulting
signal-to-noise maps. Example images of this procedure are presented
in Figure 1 of \citet{walsh2016}.

As a next step to confirm the {\sc DUCHAMP} maser detections, each one
of them was visually inspected. Almost all false detection could then
be associated with side-lobes from nearby strong masers.

In addition to the {\sc DUCHAMP} approach, we visually searched the
data cubes and moment maps using the HOPS method outlined in more
detail in \citet{walsh2012}. This was particularly useful for the
1612\,MHz masers associated with evolved stars because their typical
double-horn profile is easy to visually identify, even in low
signal-to-noise data.

To extract the parameters for each maser spot, the task {\tt imfit} in
{\sc MIRIAD} is used \citep{sault1995}. Derived parameters are the
peak and integrated flux densities, as well as the positions and their
associated uncertainties. The velocity minima and maxima for each spot
are derived via visually inspecting the data cubes over which channels
emission is observed.

The completeness of the maser detections depends on the noise level
and therefore varies spatially as well as spectrally. In spectral
channels and at positions where a strong maser spot is found nearby,
the noise level is higher, and therefore the degree of completeness
lower. Based on this, one cannot derive a single completeness level,
but that has to be inferred as a function of the changing noise
level. \citet{walsh2016} have conducted a detailed completeness
analysis for the pilot region, and we refer to their analysis. The
main conclusion is that the THOR maser data are complete at around the
level of 0.25\,Jy\,beam$^{-1}$, and that the data are still 50\%
complete at approximately 0.17\,Jy\,beam$^{-1}$.

\subsection{Maser catalogue and properties}

The full THOR OH maser catalogue consists of 1585 individual maser
spots, including the 276 spots detected in the pilot region by
\citet{walsh2016}.  The catalogue contains 1080 maser spots at
1612\,MHz, 307 at 1665\,MHz, 145 at 1667\,MHz and 53 at 1720\,MHz,
respectively. Figure \ref{samplefig1} presents an overlay of all maser
spots on the 1.4\,GHz THOR continuum image \citep{wang2018}.

The maser spots are grouped into maser sites that occur at
well-defined positions on the sky, typically smaller than $1''$ (e.g.,
\citealt{elitzur1992b}) and are always much smaller compared to the
THOR beam size between $12.5$ and $19''$ (Sect.~\ref{data}). These
maser sites can harbor of maser spots of the same line but they can
also contain maser spots from multiple transitions. Altogether the
1585 maser spots are distributed in 807 maser sites. For all maser
sites we show an 8\,$\mu$m image -- mainly from the GLIMPSE survey
\citep{benjamin2003,churchwell2009} and for the 33 sites, where no
GLIMPSE data are available, from WISE \citep{wright2010} --
accompanied by the corresponding spectra. The grey-shaded areas in
each spectrum mark the velocity regimes of each identified maser spot
also listed in Table \ref{catalogue}. Fig.~\ref{example} shows a few
examples, data for all sites are presented in the appendix.

\begin{figure}[ht]
\includegraphics[width=0.49\textwidth]{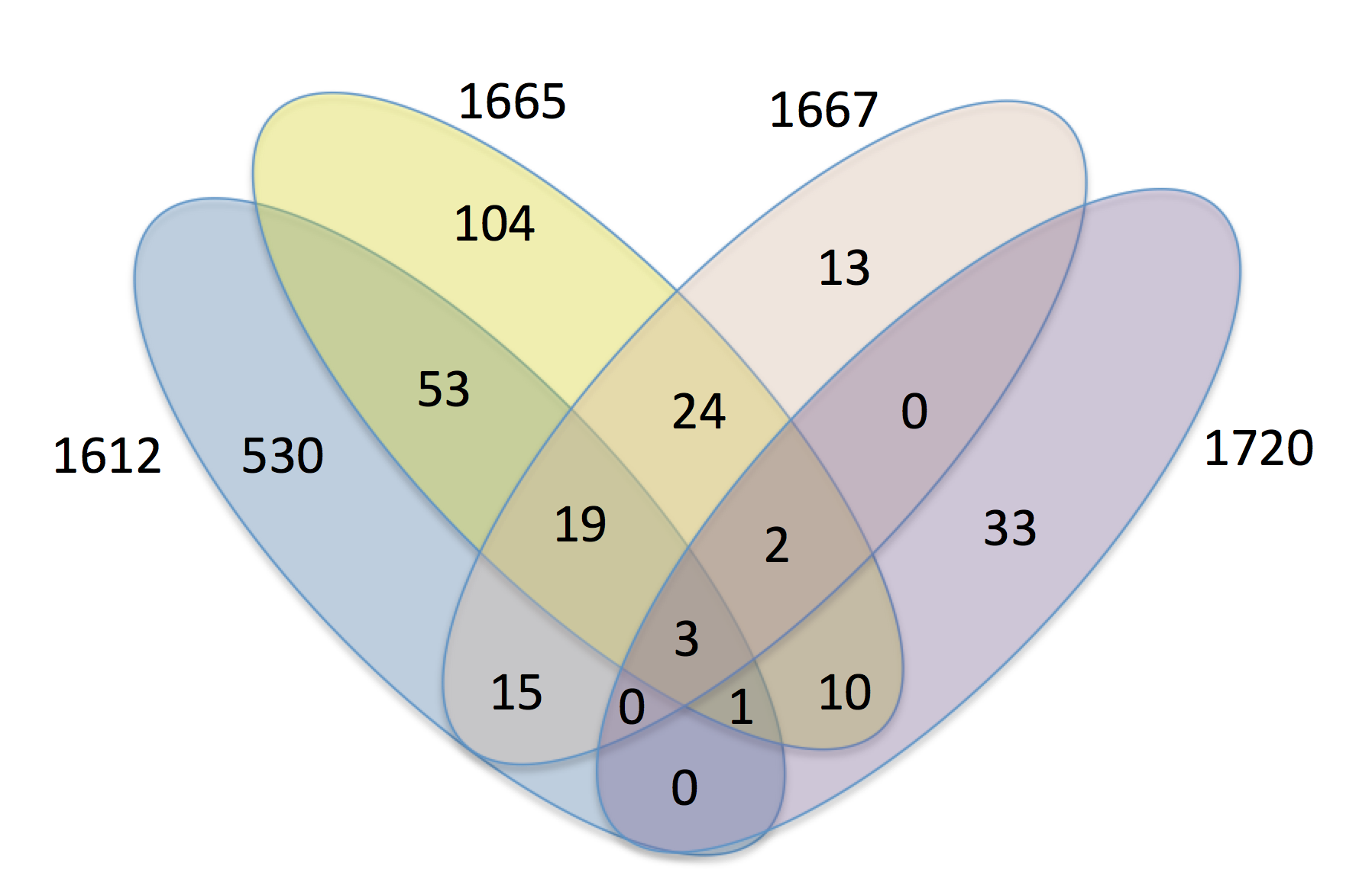}
\caption{Venn diagram showing the occurrence of masers for the
  807 OH maser sites.}
\label{venn}
\end{figure}

To evaluate the occurrence of maser spots of different OH transitions
within a single maser site, Figure \ref{venn} presents a Venn-diagram
outlining the common occurrence of maser sites for the four
transitions. The largest fraction of maser sites ($\sim$66\%) are
1612\,MHz masers without associated masers in other ground-state OH
transitions. However, in contrast to the pilot study where several of
the transition overlap areas in this diagram were not populated at
all, in the full THOR survey we find maser sites for which most
permutations of transition overlaps. Even maser sites with all four
maser types together are identified, although only three in total,
corresponding to the three sites G43.165-0.028, G43.166+0.012 (both
part of very the luminous H{\sc ii} region complex W49A), and
G45.466+0.045 which is part of the H{\sc ii} region
G45.47+0.05. Therefore, although each maser type still has a preferred
physical environment, the sole detection of an OH maser does not allow
an unambiguous identification of the underlying source type, whether
it is a star-forming region, an AGB star or a supernova remnant.

\section{Discussion}

\subsection{Maser associations}

With 807 maser sites, it is important to know the physical nature of
the underlying sources. Therefore, where possible, we have associated
the maser sites with astrophysical source types. As outlined in the
Introduction, previous studies indicate that 1612\,MHz masers are
often associated with evolved stars, 1665 and 1667\,MHz masers tend to
be associated with star-forming regions, and 1720\,MHz masers are
often associated with shocks, potentially from supernova
remnants. However, shocks could also be produced in protostellar jets
and hence maybe the 1720\,MHz masers as well. As already visualized in
the Venn diagram (Fig.~\ref{venn}), these associations are not unique,
and there are several regions where different maser types are
associated with the same astrophysical region. The overlay of the
maser sites with the cm continuum emission in Fig.~\ref{samplefig1}
already shows that only a comparatively small fraction of maser sites
is associated with radio continuum emission that traces either
free-free or synchrotron emission. At the sensitivity limits of the
THOR continuum data \citep{wang2018}, only 2.3, 19.4, 11.8 and 18.4\%
of the 1612, 1665, 1667, and 1720\,MHz OH maser sites are associated
with radio continuum emission, respectively. The corresponding 54 OH
maser sites are labeled in Table \ref{catalogue}.

The source association process was done in several steps. First, we
went through all maser spectra and mid-infrared images (Figures in
appendix), and identified those sources as evolved stars that showed
the typical double-peaked profiles in the 1612\,MHz line (ES-D in
Table \ref{catalogue}). In addition to the
clear detection associations, there are also a few 1612\,MHz masers
where the single component is close to the edge of our bandpass, and
the second component may be at velocities outside of our velocity
range. Furthermore, a few 1612\,MHz masers show a weak second
component that is below our detection threshold but may still be
real. Those sources we tentatively associated with evolved stars
(ES-D? in Table \ref{catalogue}). Furthermore, a few classical maser sources were
identified straight away, e.g., the supernova remnant W44 or the
luminous star-forming region W49. As a next step, we employed the
SIMBAD database \citep{wenger2000} to search for associations within a
$10''$ matching radius. This allowed for an additional large fraction
of source-type associations, all those are marked as ``Sim'' in Table
\ref{catalogue}. In addition to this, all
sources associated with class II methanol masers (MMB,
\citealt{green2010,breen2015}) were catalogued as star-forming
regions.

Based on these approaches, 716 of the 807 maser sites were
identified. Of these, 415 maser sites exhibit double-horned 1612\,MHz
profiles associated with the expanding shells around evolved stars, 57
sites may be double-horned and associated with evolved stars
(classified as ES-D?), 56 sites are co-spatial with stars according to
the SIMBAD database, 164 sites should be associated with star-forming
regions, 21 regions are potentially associated with supernova remnants
(SNR), 3 sites are identified as planetary nebulae (PN), and 1 is a
possible pulsar association. 90 sites remain as unidentified.

\begin{figure*}[ht]
\begin{center}
  \includegraphics[width=0.7\textwidth]{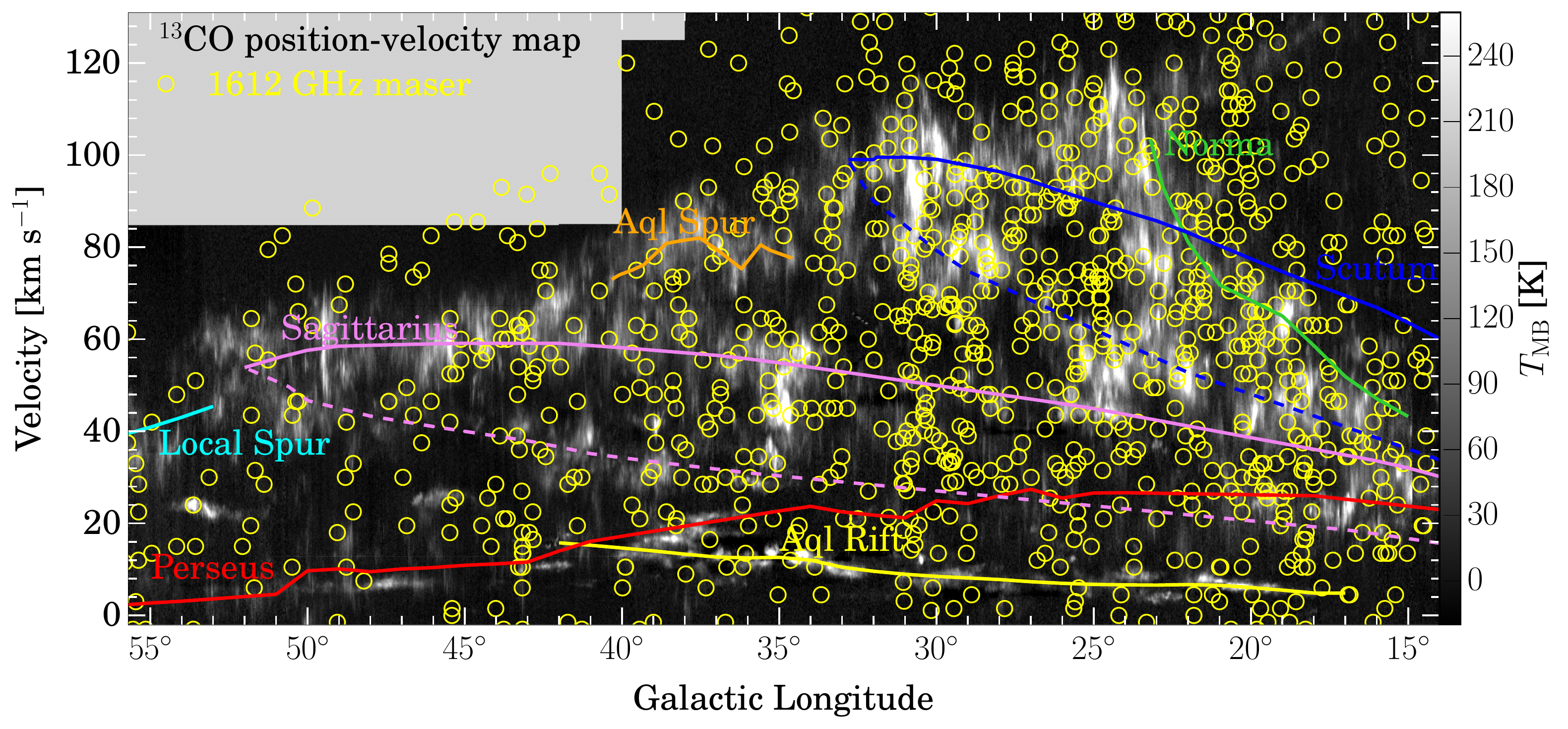}\\
  \includegraphics[width=0.7\textwidth]{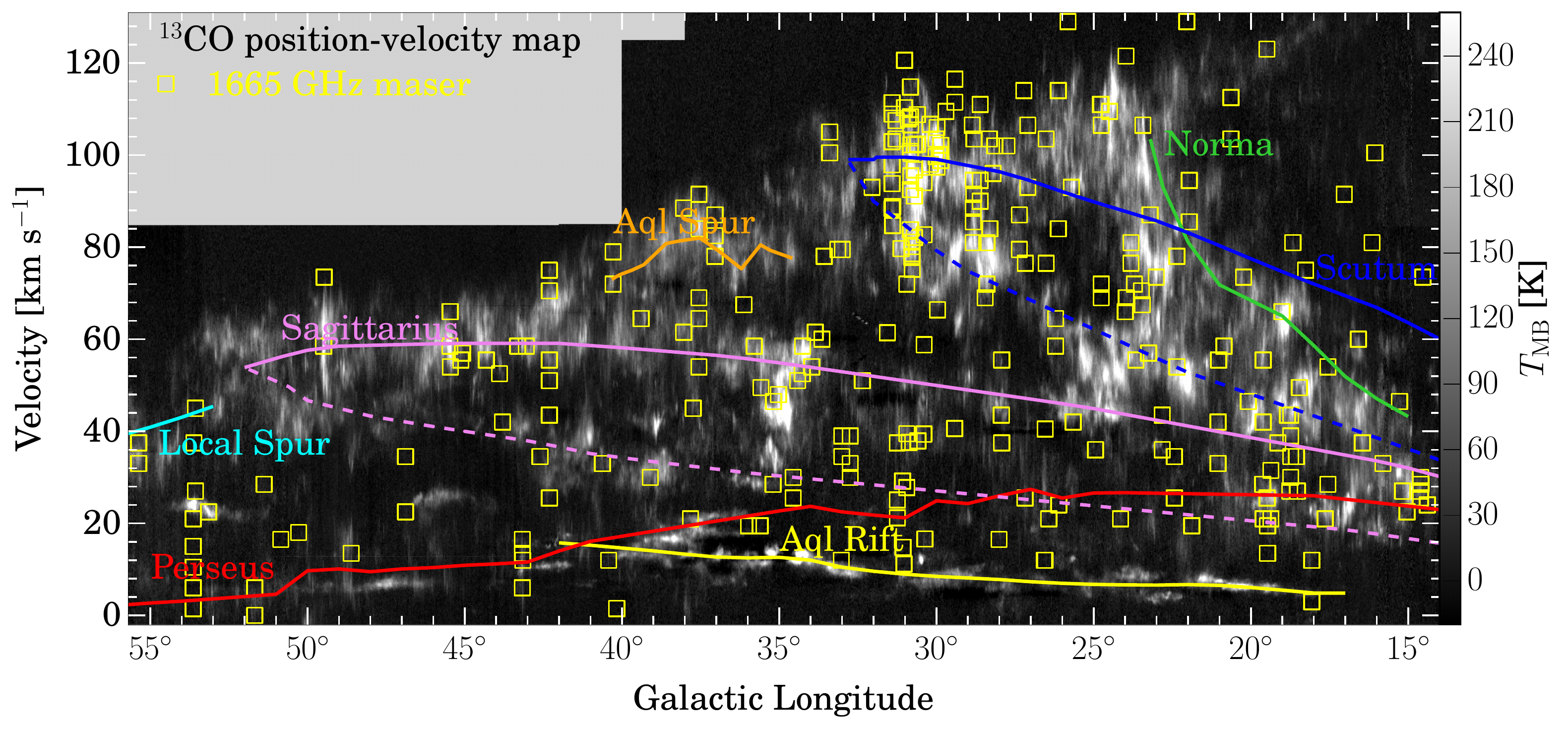}\\
  \includegraphics[width=0.7\textwidth]{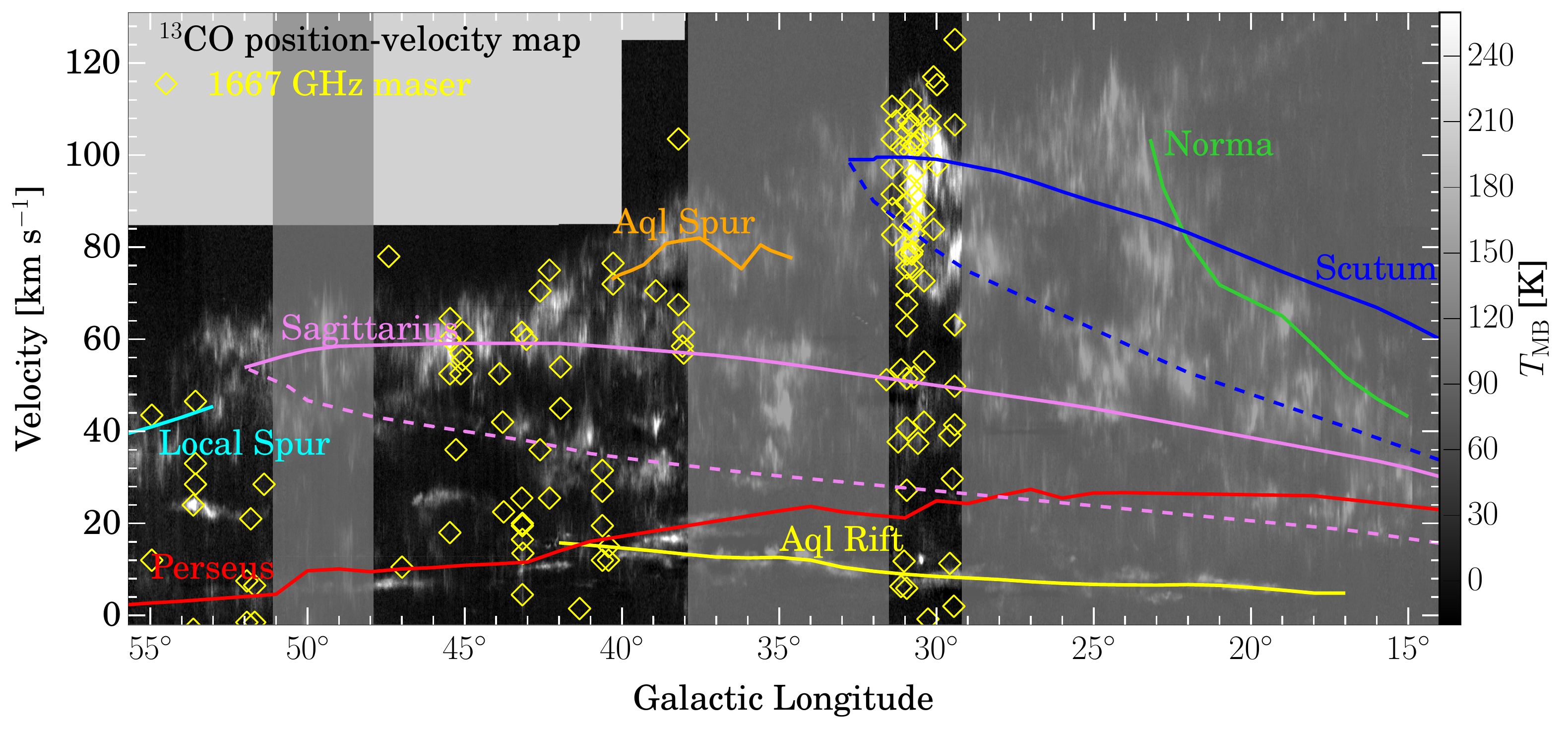}\\
  \includegraphics[width=0.7\textwidth]{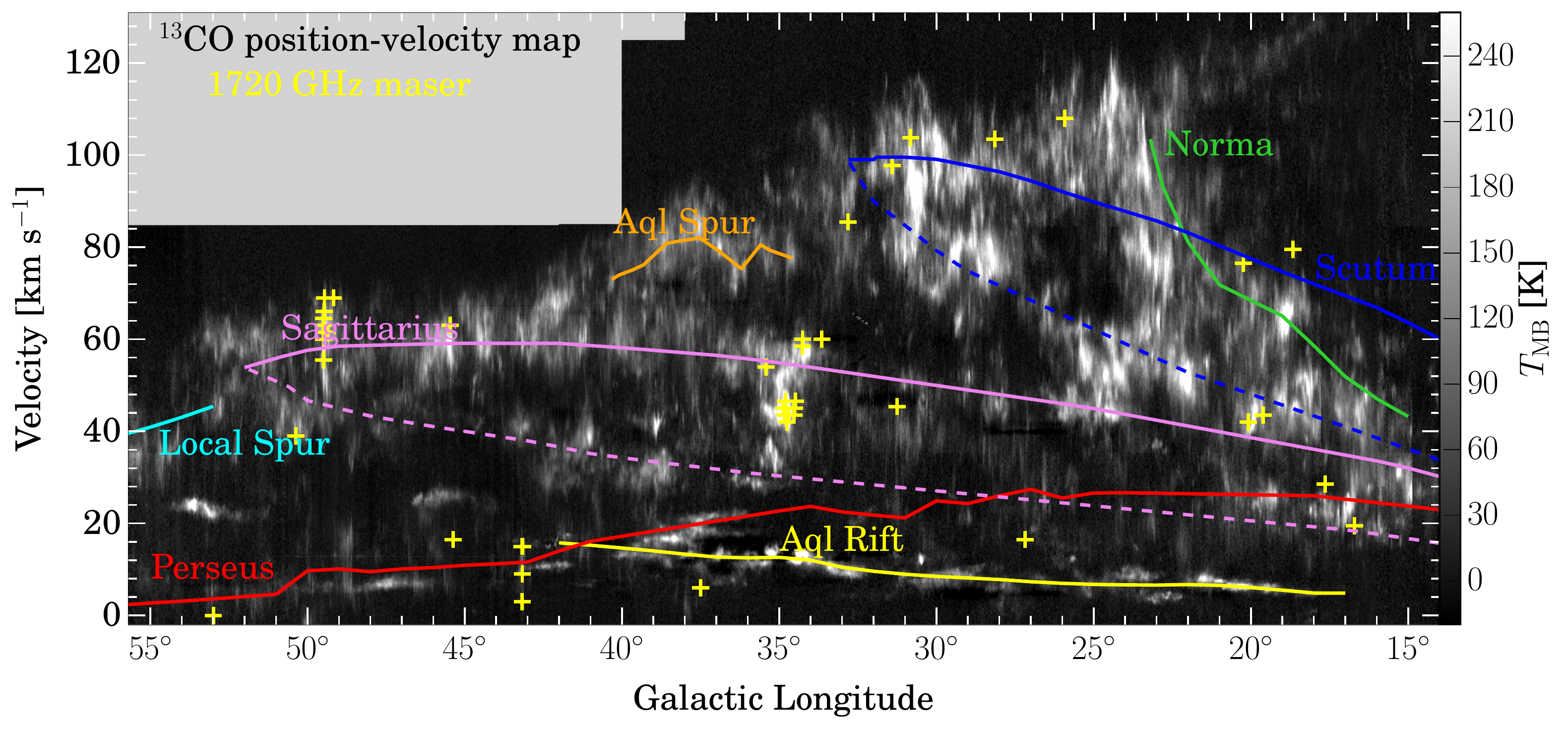}
\end{center}
\caption{Overlay of OH masers with $^{13}$CO(1--0) emission. The four
  panels show in grey-scale the position-velocity diagrams along the
  Galactic longitude axis (integrated over Galactic latitude) for the
  $^{13}$CO(1--0) data from the Galactic Ring Survey
  \citep{jackson2006}. The symbols in the four panels from top to
  bottom show the corresponding OH maser spots from the 1612, 1665,
  1667 and 1720\,MHz lines, respectively. The colored lines present
  the spiral arm models from \citet{reid2016}. The missing
    1667\,MHz masers in the grey-shaded boxes below $29\pdeg 2$, from
    $31\pdeg 5$ to $37\pdeg9$, and between $47\pdeg 1$ to $51\pdeg2$,
    are because this line was not covered in that part of the survey
    \citep{beuther2016}.}
\label{13co}
\end{figure*}

An interesting comparison is how the different OH masers compare to
the gas distribution within the Milky Way. Figure \ref{13co} shows a
position-velocity plot of the $^{13}$CO(1--0) emission along Galactic
latitudes based in the Galactic Ring Survey
\citep{jackson2006}. Over-plotted on the four panels are the different
maser types at 1612, 1665, 1667 and 1720\,MHz, respectively. One sees
clear differences between the OH 1612\,MHz masers, that are
distributed almost randomly across the figure, whereas the three other
types at 1665, 1667, and 1720\,MHz are correlated well with the
$^{13}$CO(1--0) emission in the Milky Way plane. Although the
1665/1667 and 1720\,MHz masers trace largely different physical
environments -- star formation sites versus shock excitation in
supernovae or jets (see Introduction) -- they are all closely
associated with the dense gas associated with star formation or
feedback processes. In contrast to this, the 1612\,MHz masers, that
typically trace the shells around evolved stars, do not have much in
common anymore with the original gas distributions of the birth sites.

\clearpage

\subsection{Double-horned OH 1612\,MHz separation}

\begin{figure}[ht]
\includegraphics[angle=-90,width=0.55\textwidth]{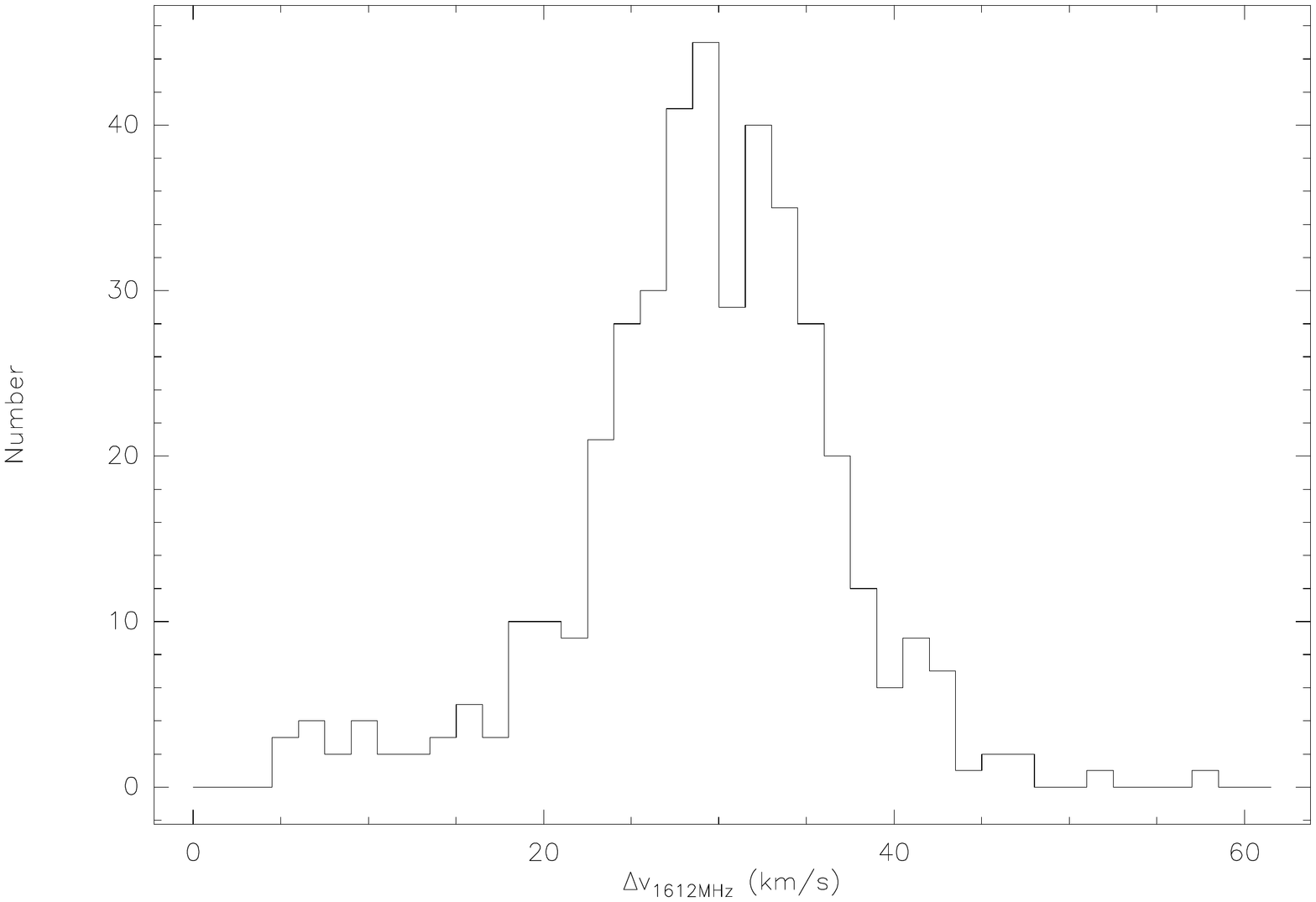}
\caption{Histogram of 1612\,MHz double-peak velocity differences.}
\label{deltav}
\end{figure}

More than 50\% of the maser sites (415) can be identified as
double-horned OH 1612\,MHz masers associated with evolved stars. The
pumping mechanism by far-infrared photons ($\sim$35\,$\mu$m) in the
expanding shells around evolved stars, that is responsible for this
spectral feature, is described in detail in \citet{elitzur1992}. Using
the peak velocities of each maser spot (Table \ref{catalogue}), we can
derive the typical velocity differences $\Delta v_{\rm 1612MHz}$
between the main double-horned profiles. Figure \ref{deltav} presents
the derived distribution. While there are a few exceptions at high and
low velocity differences, it is clear that the distribution has most
velocity separations between roughly 22 and 38\,km\,s$^{-1}$,
consistent with previous findings by \citet{baud1981} or
\citet{sevenster2001}. Fig.~\ref{deltav} exhibits two bins around
$\sim$29 and $\sim$32\,km\,s$^{-1}$ with slightly higher separation
population. However, since this is a small effect at a velocity
separation of only two times our velocity resolution, we do not
consider this further.

\subsection{1720\,MHz maser association}

Following the identification of maser sites in Table \ref{catalogue},
the rarest of the masers, those that emit in the 1720\,MHz line, are
mostly associated with supernova remnants (21 times), but a
non-negligible fraction of 17 maser sites is also associated with
star-forming regions. The remaining ones are either unidentified (6)
and 1 is associated with a planetary nebula.

The main differences between 1720\,MHz masers from supernovae remnants
and star-forming regions appears to lie in the density of the
gas. While SNR associated 1720\,MHz masers typically require
relatively low densities ($n\sim 10^4$\,cm$^{-3}$) and may trace
C-shocks (e.g.,
\citealt{pavlakis1996a,pavlakis1996b,lockett1999,caswell1999}),
1720\,MHz masers in star-forming regions likely need higher densities
($\sim 10^{6-7}$\,cm${-3}$), intense far-infrared radiation and large
velocity gradients to be effectively pumped (e.g.,
\citealt{gray1992,pavlakis1996b,caswell1999}). These different regimes
reflect that our survey reveals significant numbers of 1720\,MHz masers
toward SNRs as well as star-forming regions.

\subsection{Literature comparison}

\subsubsection{Previous VLA 1612\,MHz survey}

\begin{figure}[ht]
\includegraphics[angle=-90,width=0.72\textwidth]{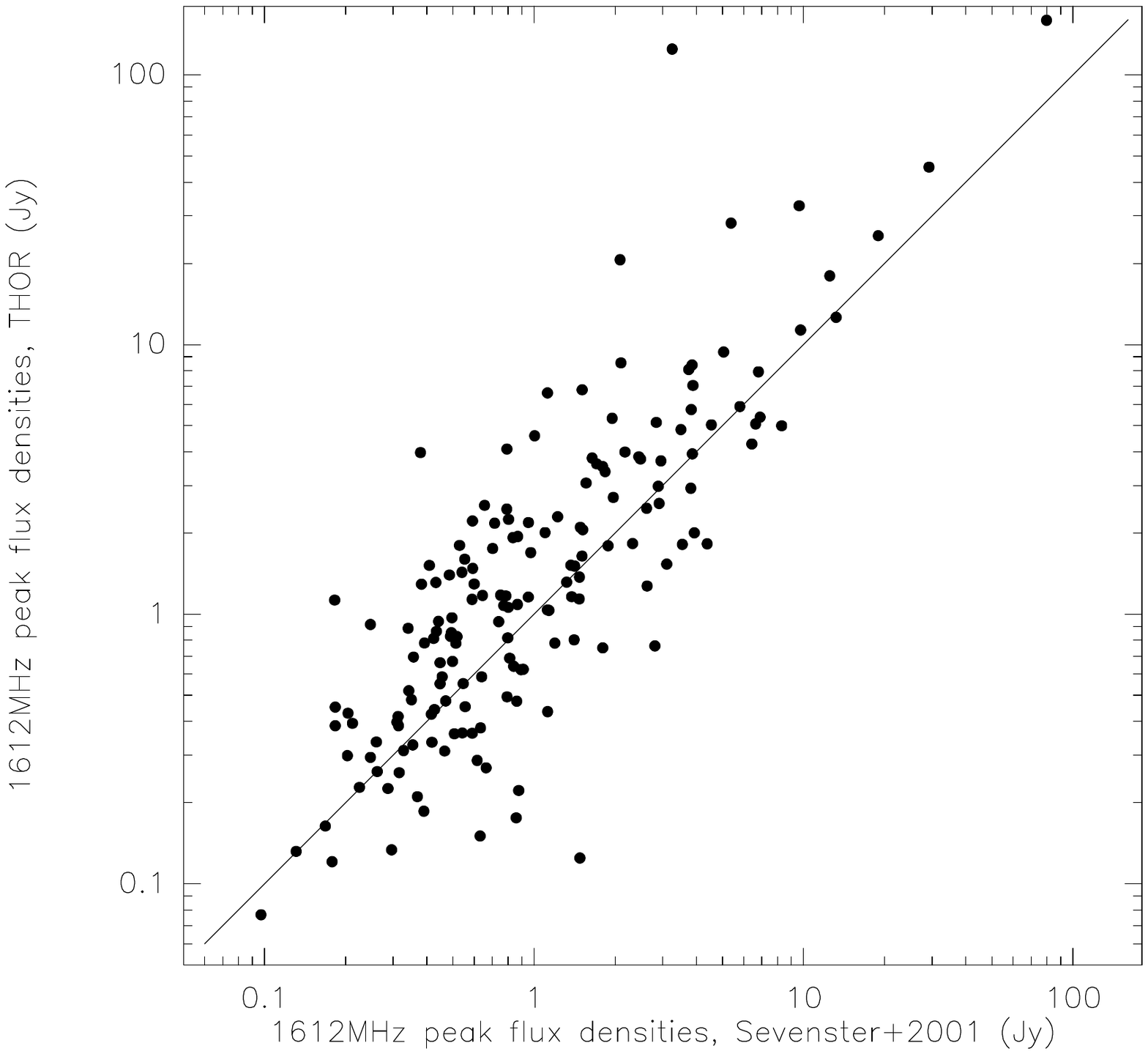}
\caption{Comparison of peak flux densities of the 1612\,MHz maser
  between the \citet{sevenster2001} survey and the new THOR study.}
\label{sevenster}
\end{figure}

As mentioned in the Introduction, \citet{sevenster2001} conducted a
VLA survey of the 1612\,MHz maser at Galactic longitudes below
45\,deg. That survey had been conducted heterogeneously, employing all
VLA configurations, and the reported median rms noise was
25\,mJy\,beam$^{-1}$ \citep{sevenster2001}. From the 1080 maser spots
identified at 1612\,MHz, the \citet{sevenster2001} survey had
previously identified 166 sources in the THOR survey area with a
matching radius of $10''$. Hence, most of the THOR-identified masers
are new detections. Fig.~\ref{sevenster} presents a comparison between
the peak flux densities measured by the earlier \citet{sevenster2001}
study and our new THOR data. Although there is a large scatter, in the
log-log plot both surveys correlate relatively well with each
other. However, if one measures the ratio of the mean THOR flux
densities over the mean flux densities from \citet{sevenster2001}, one
finds systematically higher THOR values with a mean ratio of
$\sim$1.9. Similarly, previous studies by \citet{dawson2014} comparing
the SPLASH survey with the \citet{sevenster2001} study, and the THOR
pilot study \citep{walsh2016} found lower flux densities in the
\citet{sevenster2001} study. Following their arguments, the most
likely explanation for this systematic difference is the broader
channel width of 2.27\,km\,s$^{-1}$ used by \citet{sevenster2001} that
can smear out the intrinsically higher peak flux densities.

\subsubsection{Parkes 1665 and 1667\,MHz survey}

\begin{figure}[ht]
\includegraphics[angle=-90,width=0.72\textwidth]{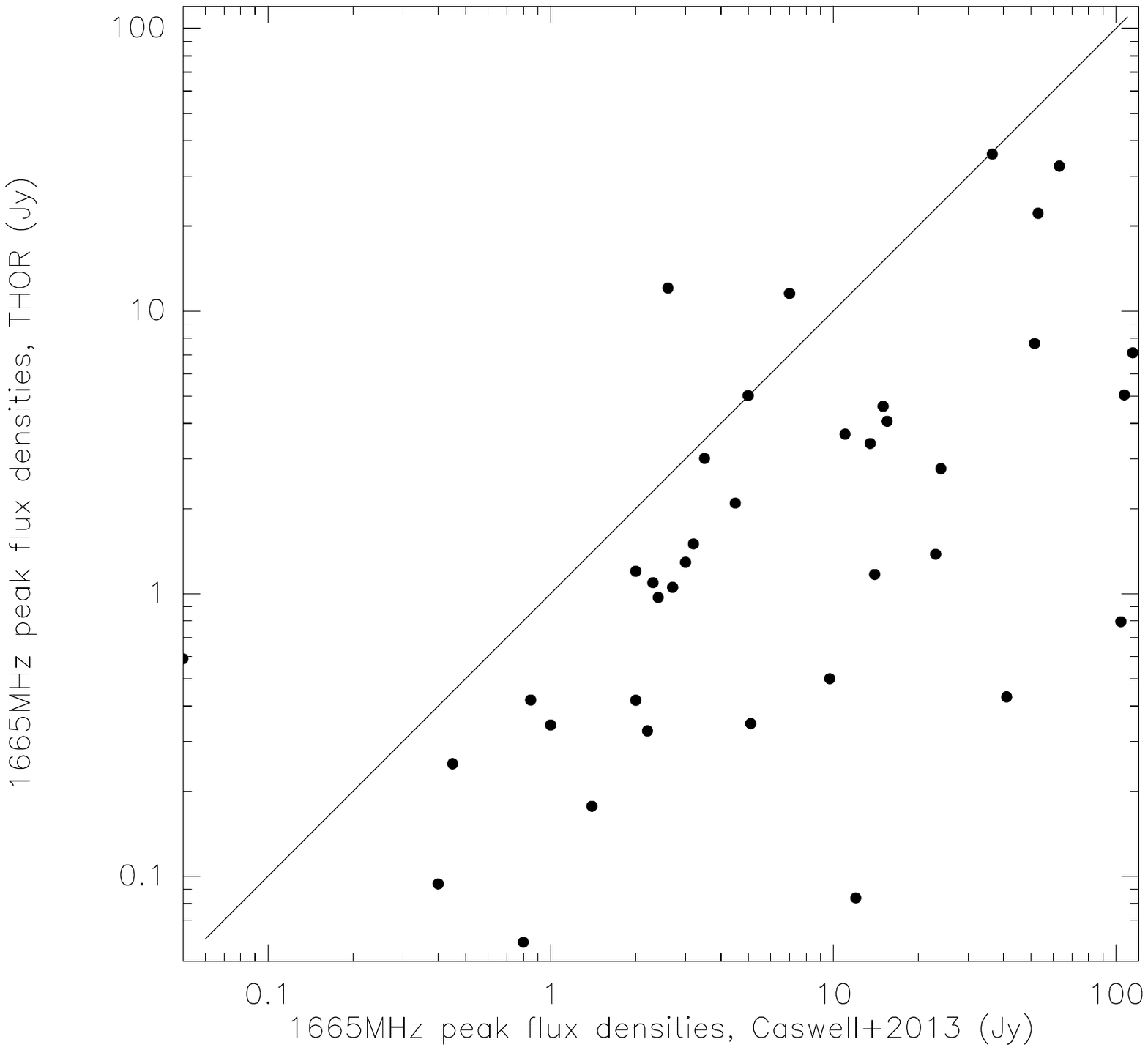}
\caption{Comparison of peak flux densities of the 1665\,MHz maser
  between the \citet{caswell2013} survey and the new THOR study.}
\label{caswell}
\end{figure}

\citet{caswell2013} presented main-line (1665 and 1667\,MHz) Parkes
single-dish maser observations toward 104 sources below Galactic
longitudes of 41\,deg. Comparing the 1665\,MHz detections with the
THOR survey (again within a positional matching radius of $10''$),
both have 38 1665\,MHz maser spots in common. Fig.~\ref{caswell} shows
a comparison between the peak flux densities of the
\citet{caswell2013} and the THOR data, and interestingly the
Parkes-measured flux densities are consistently much higher than what
we observed with the VLA. Following the arguments above for the
1612\,MHz maser comparison, also here the spectral resolution is
likely to be the explanation. Compared to the THOR velocity resolution
of 1.5\,km\,s$^{-1}$, the Parkes data were observed with a much better
spectral resolution of $\sim$0.1\,km\,s$^{-1}$. Hence, for the
1665\,MHz comparison, the argument is similar, i.e., the
higher-spectral-resolution Parkes data can measure higher peak flux
densities that are smeared out in our lower-spectral-resolution THOR
data. In addition to this, the Parkes observations with a beam size of
$\sim$12$'$ may add several maser sites into a single spectrum that we
can resolve with our higher spatial resolution, hence finding lower
flux densities in individually resolved sites.

\subsubsection{Class II CH$_3$OH masers}

Other maser associations interesting for comparison are the class II
CH$_3$OH masers typically associated with young high-mass star-forming
regions (e.g., \citealt{walsh1998}). Comparisons between OH
  and class II CH$_3$OH maser sites with ultracompact H{\sc ii}
  regions indicated that OH masers are more often associated with the
  ionized gas and hence most likely trace a later evolutionary stage
  \citep{caswell1997}. Similar relative evolutionary stages were more
  recently reported for OH and 6.7\,GHz class II masers by
  \citet{breen2010,breen2018}. The Methanol Multibeam survey (MMB)
searched with the Parkes telescope for these CH$_3$OH masers, and they
provide a catalogue between Galactic longitudes of 345 and 60\,degrees
\citep{caswell2010,green2010,breen2015}. In the overlap region with a
matching radius of $10''$, 108 maser sites coincide between our
  THOR and the MMB surveys. Since class II CH$_3$OH masers should be
associated with high-mass star-forming regions, indeed most MMB sites
are associated with the 1665\,MHz OH masers (101 sites). Because the
real coverage of the 1667\,MHz masers is smaller (Section
\ref{data}), only 23 are associated with the MMB sources. From the
rarer 1720\,MHz masers, 13 have a MMB counterpart and hence should be
associated with star formation. Furthermore, as expected the 1612\,MHz
masers are only rarely associated with CH$_3$OH class II masers, and
we find only 15 sites in common. Nevertheless, these should be
star-forming regions as well. Most of the 1612, 1667 and 1720 masers
have also a counterpart in the 1665\,MHz line, more details about the
maser associations are given in Table \ref{masers}.

Considering relative time-scales when these masers occur during
high-mass star formation, one can compare the occurrence of class II
CH$_3$OH masers to that of the 1665\,MHz OH masers. In the common area
up to Galactic longitudes of 60\,deg, there are 319 MMB maser sites
and 215 OH 1665\,MHz maser sites. Of these, 101 sites are in
common. Hence, one finds a similar number of 114 OH 1665\,MHz masers
without MMB counterpart, and 218 MMB sites without 1665\,MHz
counterpart. Therefore, relatively speaking, the time-scale to find
6.7\,GHz class II CH$_3$OH maser may be approximately 50\% longer than
that for the 1665\,MHz OH maser. \citet{breen2018} compared the
occurrence of the 6.7\,GHz class II CH$_3$OH masers with the excited
OH masers at 6035\,MHz, and they also find that around 54\% of these
excited OH masers are associated with class II CH$_3$OH masers. They
report that their results are consistent with previous findings that
the OH masers tracing a slightly more evolved evolutionary stage than
the class II CH$_3$OH masers (e.g.,
\citealt{caswell1997,breen2010,breen2018}).

\begin{table}[htb]
  \caption{Numbers of associated maser sites}
  \label{masers}
  \begin{tabular}{l|r|r}
    \hline
    \hline
    & CH$_3$OH & H$_2$O \\
    \hline
1612& 15 (2)   & 20 (9) \\
1665& 101      & 33     \\
1667& 23 (1)   & 2 (2)  \\
1720& 13 (3)   & 4 (0)  \\
  \hline
  \hline
  \end{tabular}
  ~\\
  Notes: The numbers in brackets state how many of the maser sites
  have no 1665\,MHz counterpart.
\end{table}
    
\subsubsection{H$_2$O masers}

Although the spatial overlap is smaller, we also compared our OH maser
catalogue to the H$_2$O maser catalogue HOPS presented in
\citet{walsh2011}. In the overlap region up to 30\,deg in
longitude, the HOPS catalogue lists 571 H$_2$O maser sites. In total,
the THOR and HOPS catalogues do have only 44 maser sites in common
(more details in Table \ref{masers}). The most common association (33)
is again that with the 1665\,MHz maser. Most of these (25) are
associated with star-forming regions (20 with MMB counterpart and 5
classified as SF from SIMBAD). The remaining source associations for
these masers are 2 evolved stars (ES), 3 SIMBAD-identified Stars, and
3 unidentified sources.

With the narrow spatial overlap of the 1667\,MHz masers, only 2 of
them have a HOPS counterpart, both are unidentified sources and none
have a 1665\,MHz counterpart.

Interestingly, also the 1720\,MHz masers have 4 associated HOPS
sources, all related to star formation (3 MMB sources and 1
SIMBAD-identified star-forming region).

Finally, 20 regions with 1612\,MHz maser emission are also associated
with HOPS H$_2$O masers. Of these 20 regions, 13 are associated with
(evolved) stars (10 ES and 1 SIMBAD-identified star in our
catalogue). A further 6 regions are associated with star formation
(MMB sources) and 1 remains an unidentified source.

\section{Conclusions and Summary}
\label{conclusion}

Based on the HI/OH/Recombination line survey of the Milky Way (THOR),
we present an OH maser catalogue in the longitude/latitude ranges
$14.3<l<66.8$ and $b<\pm 1.25$\,degrees. All ground state OH lines at
1612, 1665, 1667 and 1720\,MHz are covered. In total, we identify 1585
individual maser spots distributed over 807 maser sites. Where
possible, associations with astrophysical source types are conveyed
(evolved stars, star-forming regions, supernovae remnants). More than
50\% of the maser sites are associated with the double-horned
1612\,MHz profiles typically stemming from the expanding shells of
evolved stars. A comparison with the Galactic $^{13}$CO(1--0) emission
as well as other maser surveys (OH as well as class II CH$_3$OH and
H$_2$O masers) is presented. The full catalogue is available in
electronic form.

\begin{acknowledgements} 
  We thank an anonymous referee for careful comments improving this
  paper. The National Radio Astronomy Observatory is a facility of the
  National Science Foundation operated under cooperative agreement by
  Associated Universities, Inc. HB, YW, JS and MR acknowledge support
  from the European Research Council under the Horizon 2020 Framework
  Program via the ERC Consolidator Grant CSF-648505. HB, MR, SCOG and
  RSK acknowledge support from the Deutsche Forschungsgemeinschaft via
  SFB 881, “The Milky Way System” (sub-projects B1, B2 and B8). JK
  received funding from the European Union’s Horizon 2020 research and
  innovation program under grant agreement No 639459 (PROMISE). NR
  acknowledges support from the Infosys Foundation through the Infosys
  Young Investigator grant. F.B. acknowledges funding from the
  European Union’s Horizon 2020 research and innovation programme
  (grant agreement No 726384). This research has made use of the
  SIMBAD database, operated at CDS, Strasbourg, France.
\end{acknowledgements}


\end{document}